\documentclass[a4paper,11pt]{article}

\usepackage{jheppub}
\usepackage{lmodern} 
\usepackage[T1]{fontenc} 
\usepackage{microtype} 

\title{Hamilton-Jacobi approach to holographic renormalization of massive gravity}

\author[a]{Fan Chen,}
\author[a,b]{Shao-Feng Wu}
\author[c]{and Yuxuan Peng}

\affiliation[a]{Department of Physics, Shanghai University, Shanghai, 200444, China}
\affiliation[b]{Center for Gravitation and Cosmology, Yangzhou University, Yangzhou, 225009, China}
\affiliation[c]{CAS Key Laboratory of Theoretical Physics, Institute
	of Theoretical Physics, \\Chinese Academy of Sciences, Beijing, 100190, China}

\emailAdd{fanchen@shu.edu.cn}
\emailAdd{sfwu@shu.edu.cn}
\emailAdd{yxpeng@itp.ac.cn}

\abstract{Recently, a practical approach to holographic renormalization has been
	developed based on the Hamilton-Jacobi formulation. Using a simple
	Einstein-scalar theory, we clarify that this approach does not conflict with
	the Hamiltonian constraint as it seems. Then we apply it to the holographic
	renormalization of massive gravity. We assume that the shift vector is falling off fast enough asymptotically. We derive the counterterms up to the
	boundary dimension \emph{d}=4. Interestingly, we find that the conformal anomaly can even occur in odd dimensions, which is different from the Einstein gravity.
    We check that the counterterms cancel the divergent part of the on-shell action at the background level. At the perturbation level, they are also applicable in several time-dependent cases.}


\arxivnumber{1903.02672}

\begin{document}
\maketitle
\flushbottom

\section{Introduction}

Anti-de Sitter/conformal field theory (AdS/CFT) correspondence provides a
powerful tool to study the strongly coupled field theories \cite{Aharony9905}%
. Among others, the Gubser--Klebanov--Polyakov--Witten dictionary that
identifies the generating functional of the field theory with the on-shell
gravitational action plays an essential role in the calculation \cite%
{Gubser0112,Witten9802}. The most obvious technical obstacle to use the
dictionary is the divergence involved on both sides of the duality \cite%
{Papadimitriou2016}. According to the renormalization method to deal with
the UV divergence in the field theory, the called holographic
renormalization is developed to remove the IR divergence in the gravity.

There are different approaches to holographic renormalization. The first
systematic one was presented in \cite{Henningson9806,deHaro0002,Bianchi0112}%
, which is usually called as the standard approach \cite{Papadimitriou0404}.
Its main procedure includes: a) solving the second-order equations of motion
(EOM) in the Fefferman-Graham (FG) coordinates to obtain the asymptotic
expansion of the dynamical fields \cite{Fefferman1985}; b) calculating the
regularized on-shell action on the boundary to separate the divergent terms;
c) reversing the FG expansion to express the divergent terms by the local
fields on the boundary. The standard approach is strict, conceptually
simple, and universal for diverse situations. However, the FG expansion and
its reverse are technically tedious. So it is natural to expect an
alternative approach which always respects the local field expression.

Actually, such approach was put forward by de Boer, Verlinde, and Verlinde
(dBVV) based on the Hamiltonian formulation of gravity \cite{deBoer9912},
see ref. \cite{deBoer0101} for a nice review. To proceed in dBVV's approach,
one writes down the most general ansatz for the covariant counterterms,
organizes it by the derivative expansion, and specifies it by solving a
series of descent equations induced from the Hamiltonian constraint, where
the canonical momenta are replaced by the variations of the on-shell action
with respect to boundary fields. Comparing the standard and dBVV's
approaches, one can find that the latter is usually more simple than the
former, mainly because the latter solves the algebraic descent equations
instead of the second-order differential equations, and determines the
counterterms directly on the cutoff surface without performing the FG
expansion and reversion. The main drawbacks of dBVV's approach are \cite%
{Martelli0205,Papadimitriou2016}: a) the solutions of some descent equations
are not unique; b) the logarithmic counterterms have not been explicitly
obtained; c) the ansatz may include many unnecessary terms; d) sometimes the
sufficient ansatz is difficult to be figured out. In refs. \cite%
{Kalkkinen0103,Martelli0205}, Kalkkinen, Martelli and Muck removed the
ambiguities in the descent equations by comparison with free field
calculations. They also isolated the logarithmic counterterms, which are
related to the breakdown of the recursion of descent equations.
Subsequently, Papadimitriou and Skenderis \cite%
{Papadimitriou0404,Papadimitriou0407} developed the previous approaches
where the crucial difference is that the covariant expansion is organized
according to the eigenvalues of the dilatation operator. Interestingly, this
approach does not rely on the ansatz and can be applied to more general
backgrounds \cite{Papadimitriou1106,Chemissany1405,Chemissany1408}.

Besides the standard and Hamiltonian approaches, Brown and York in the early
days proposed to remove the divergence of the stress tensor by subtracting
the contribution from the reference spacetime \cite{Brown1993}. This
requires that a boundary with intrinsic metric is embedded in the reference
spacetime, which is often not possible \cite{Balasubramanian9902}. Moreover,
instead of selecting the Dirichlet boundary conditions, the Kounterterm
approach is developed where the variational principle is associated with the
fixed extrinsic curvature on the boundary \cite{Olea0504,Olea0610}. Other
attempt based the dimensional renormalization can be found in \cite%
{Bzowski1612}.

As explicitly pointed out in dBVV's work \cite{deBoer9912}, the Hamiltonian
constraint ensures the invariance under the diffeomorphism along the radial
direction. This implies that the on-shell action does not depend on the
radial coordinate explicitly and the radial Hamilton-Jacobi (HJ) equation is
equivalent to the Hamiltonian constraint for any holographic theories with
diffeomorphism symmetry \cite{Papadimitriou1106}. On the contrary, by
focusing on the complete HJ equation rather than the formally simpler
Hamiltonian constraint, a new approach to holographic renormalization has
been presented recently \cite{Elvang1603}. This approach, which we will
refer as the HJ approach,\footnote{%
To be clear, we have referred the previous approaches based on the
Hamiltonian constraint as the Hamiltonian approaches, %
following \cite{Papadimitriou0404}.
However, it should be stressed that the Hamiltonian
approaches also solve the HJ equation.} is partially motivated by \cite%
{Larsen0307}, where the interesting point captured by \cite{Elvang1603} is
that the HJ equation is used to isolate the infrared divergences of scalar
fields in a fixed de Sitter background.
Although the HJ approach suffers from the latter two drawbacks of dBVV's approach since the action
ansatz is still required, it has been exhibited in several Einstein-scalar
theories that the HJ approach is practical \cite{Elvang1603}. Here we
emphasize that it is tailored to handle the systems with conformal
anomalies, because the derivations of the logarithmic and power counterterms
are equivalently fluent and have nothing different such as the breakdown of
descent equations. However, the reason why the HJ approach does not conflict
with the Hamiltonian constraint has not been clarified.\footnote{%
It was argued in \cite{Elvang1603} that the on-shell action is not
diffeomorphism-invariant along the radial direction and the HJ equation
cannot be reduced to the Hamiltonian constraint. Moreover, the discussion
below their eq. (2.8) might suggest that the canonical momenta in the
Hamiltonian constraint are not equal to the ones in the HJ equation.%
} In this paper, one of two aims is to address this problem.

Another aim of this paper is to apply the HJ approach to the massive gravity
with different dimensions. The research on massive gravity has a long
history \cite{Pauli1939,Veltman1970,Zakharov1970,Vainshtein1972,Boulware1972}%
. The two main motivations include finding a self-consistent theory with
massive spin-2 graviton and modifying the Einstein gravity at long distance
for self-accelerated expansion of the Universe \cite{Hinterbichler1105}.
Massive gravity has obtained revived interest since de Rham, Gabadadze, and
Tolley (dRGT) proposed a covariant non-linear theory where the\ well-known
Boulware-Deser ghost can be excluded \cite{deRham2010,deRham2011,deRham1401}%
. Recently, massive gravity has been applied to the AdS/CFT correspondence,
where the reference metric can imitate the mean-field disorder in realistic
materials \cite{Vegh1301,Davison2013,Blake2013,Blake2014,Baggioli2014}. The
holographic renormalization of massive gravity with boundary dimension $d=3$
has been studied previously using the standard approach \cite{Cao1509}.
However, the resultant counterterms are not general,
because the Gauss
normal coordinate (GNC) is adapted in the neighborhood of the boundary and some additional conditions are
imposed on the characteristic tensor of massive gravity. In this paper, we
will only assume that the GNC is applicable near the boundary but release the other conditions. Moreover, we will show that the conformal
anomalies can occur in both odd and even dimensions, which are missed in
\cite{Cao1509}. As we have emphasized, this indicates that the HJ approach
is particularly suitable for massive gravity.

The rest part of this paper is arranged as follows. In Section 2, we will
decompose the HJ equation and construct an equation that is actually used by
holographic renormalization. In Section 3, we will apply the HJ approach to
the massive gravity with different dimensions. The conclusion of this paper
will be given in Section 4. In Appendix A, we will review the HJ approach to
the holographic renormalization of the Einstein gravity with massive
scalars. In Appendix B and C, we will provide some calculation details and
basic formulas.

\section{Decomposition of Hamilton-Jacobi equation}

The bulk dynamics of a holographic theory can be formulated as a Hamiltonian
system, where the Hamiltonian time is identified with the radial coordinate $%
r$. The Hamiltonian and on-shell action still obey the HJ equation%
\begin{equation}
H+\frac{\partial S_{\mathrm{on-shell}}}{\partial r}=0,  \label{HJ0}
\end{equation}%
see a simple derivation in \cite{Papadimitriou1106}. However, one should be
careful that the diffeomorphism symmetry, which is respected by usual
gravity theories, imposes the Hamiltonian constraint $H=0$. It further
indicates that the on-shell action does not depend on $r$ explicity.
Moreover, since the Hamiltonian constraint is a part of EOM, the on-shell
action cannot be well-defined before imposing the Hamiltonian constraint.
Keeping these in mind, the Hamiltonian constraint is usually understood as
the HJ equation in the previous Hamiltonian approaches.

In ref. \cite{Elvang1603}, the Hamiltonian constraint is not imposed at the
beginning as usual. Instead, the complete HJ equation is relied on. Then the
coefficients in the action ansatz are allowed to depend on the radial
coordinate and the HJ equation induces the one-order differential equations
of the coefficients which can be solved unambiguously near the boundary. One
can find that this approach to the holographic renormalization is practical
indeed but its legitimacy has not been clearly stated. Here we will address
this problem.

Suppose that there is a general gravity theory associated with certain terms
in the action which break the diffeomorphism symmetry. Its Hamiltonian can
be nonvanishing, just like the massive gravity \cite{deRham2010,deRham2011,deRham1401}. But the HJ equation should still hold,
if the theory is still a Hamiltonian system. Turning off the
symmetry-breaking terms, one can see that $H=0$ and $\partial S_{\mathrm{on-shell}}/\partial r=0$ arise. However, the HJ equation (\ref{HJ0}) itself is not
wrong, at least formally. Thus, we can argue that the HJ equation is a more
general equation than the Hamiltonian constraint and can be applicable to
the theories with or without the diffeomorphism symmetry.

We proceed to separate the on-shell action into the renormalized part and
the divergent part%
\begin{equation}
S_{\mathrm{on-shell}}=S_{\mathrm{ren}}-S_{%
\mathrm{ct}}, \label{S2}
\end{equation}%
where the divergent terms are denoted as
negative counterterms. Then the HJ equation can be decomposed into%
\begin{equation}
H_{\mathrm{ren}}+\frac{\partial
S_{\mathrm{ren}}}{\partial r}-H_{\mathrm{ct}}-\frac{\partial
S_{\mathrm{ct}}}{\partial r}=0,  \label{HJ2}
\end{equation}%
where $H_{\mathrm{ren}}$ is defined as the part of $H$ relevant to $S_{%
\mathrm{ren}}$ and
 $H_{\mathrm{ct}}$ is
defined as%
\begin{equation}
H_{\mathrm{ct}}\equiv -\left(
H-H_{\mathrm{ren}}\right) .  \label{H2}
\end{equation}%
We point out that what is actually used to implement the holographic
renormalization of the Einstein-scalar theories in \cite{Elvang1603} is%
\footnote{%
As an illustration, we recover the holographic renormalization of the
Einstein gravity with massive scalars based on this equation in Appendix A.}
\begin{equation}
H_{\mathrm{ct}}+\frac{\partial
S_{\mathrm{ct}}}{\partial r}=0.  \label{HJdiv}
\end{equation}%
We emphasize that each term in eq. (\ref{HJ2}) should include the finite
terms if there are conformal anomalies. This subtlety implies that eq. (\ref%
{HJdiv}) is not simply the leading orders of eq. (\ref{HJ0}). Therefore,
whether it is correct or not requires proof. In the following, we will
illustrate eq. (\ref{HJdiv}) using the Einstein gravity with massive
scalars.
In particular, the Hamiltonian constraint $H=0$ will not be involved explicitly.
We argue that the extension to other theories, with or without the diffeomorphism symmetry, should be straightforward.
Note that for convenience, we will refer eq. (\ref{HJdiv}) as the counterterm part of the HJ (CPHJ) equation.

Consider that the system is described by the action%
\begin{equation}
S=-\frac{1}{2\kappa ^{2}}\int_{M}d^{d+1}x\sqrt{g}\left( R[g]-g^{\mu \nu
}G_{IJ}\partial _{\mu }\Phi ^{I}\partial _{\nu }\Phi ^{J}-V(\Phi )\right) -%
\frac{1}{\kappa ^{2}}\int_{\partial M}d^{d}x\sqrt{\gamma }K,  \label{action}
\end{equation}%
where $G_{IJ}$ is a metric on the scalar manifold, $g_{\mu \nu }$ is the
bulk metric, $\gamma _{ij}$ is the metric on the boundary, and $K$ is its
extrinsic curvature. Adopting the Arnowitt-Deser-Misner (ADM) decomposition%
\footnote{%
We denote the bulk and boundary coordinates by Greek and Latin indices,
respectively. Throughout this paper we take the Euclidean signature and set
the AdS radius $l=1$.}%
\begin{equation}
ds^{2}=g_{\mu \nu }dx^{\mu }dx^{\nu
}=(N^{2}+N_{i}N^{i})dr^{2}+2N_{i}drdx^{i}+\gamma _{ij}dx^{i}dx^{j},
\label{ADM}
\end{equation}%
and selecting the usual gauge due to the diffeomorphism symmetry%
\begin{equation}
N=1,\qquad N_{i}=0,  \label{gauge0}
\end{equation}%
where $N$ is the lapse and $N^{i}$ is the shift, the Hamiltonian is given by%
\footnote{%
In Appendix A.1, we have reviewed briefly the Hamiltonian formalism of the
Einstein-scalar theory.}
\begin{equation}
H=\int_{\partial M}d^{d}x\left[ \frac{2\kappa ^{2}}{\sqrt{\gamma }}\left(
\pi _{ij}\pi ^{ij}-\frac{1}{d-1}\pi ^{2}+\frac{1}{4}G^{IJ}\pi _{I}\pi
_{J}\right) +L_{d}\right] ,
\end{equation}%
where%
\begin{equation}
L_{d}=\frac{\sqrt{\gamma }}{2\kappa ^{2}}\left( R-\gamma ^{ij}G_{IJ}\partial
_{i}\Phi ^{I}\partial _{j}\Phi ^{J}-V(\Phi )\right) ,
\end{equation}%
and $R$ is the scalar curvature on the boundary. The canonical momenta are
defined by%
\begin{eqnarray}
\pi ^{ij}\equiv \frac{\partial L}{\partial \dot{\gamma}_{ij}} &=&\frac{1}{%
2\kappa ^{2}}\sqrt{\gamma }(K^{ij}-K\gamma ^{ij}),  \nonumber \\
\pi _{I}\equiv \frac{\partial L}{\partial \dot{\Phi}^{I}} &=&\frac{1}{\kappa
^{2}N}\sqrt{\gamma }(G_{IJ}\dot{\Phi}^{J}-N^{i}G_{IJ}\partial _{i}\Phi ^{J}).
\label{CM D1}
\end{eqnarray}%
According to the standard classical mechanics \cite{Landau1987}, they should
be equal to the variations of the on-shell action with respect to boundary
fields%
\begin{equation}
\pi ^{ij}=\frac{\delta S_{\mathrm{on-shell}}}{\delta \gamma _{ij}},\qquad
\pi _{I}=\frac{\delta S_{\mathrm{on-shell}}}{\delta \Phi ^{I}}.
\label{momenta2}
\end{equation}%
Using eq. (\ref{momenta2}), the previous decomposition of the HJ equation
indicates%
\begin{eqnarray}
H_{\mathrm{ren}} &=&\int_{\partial M}d^{d}x\left[ 2\{-S_{\mathrm{ct}},S_{%
\mathrm{ren}}\}+\{S_{\mathrm{ren}},S_{\mathrm{ren}}\}\right] ,  \label{Hren}
\\
H_{\mathrm{ct}} &=&-\int_{\partial M}d^{d}x\left[ \{S_{\mathrm{ct}},S_{%
\mathrm{ct}}\}+L_{d}\right] ,  \label{Hnonren}
\end{eqnarray}%
and the bracket $\{S_{\mathrm{a}},S_{\mathrm{b}}\}$ is defined through%
\begin{equation}
\{S_{\mathrm{a}},S_{\mathrm{b}}\}\equiv \frac{2\kappa ^{2}}{\sqrt{\gamma }}%
\left( \frac{\delta S_{\mathrm{a}}}{\delta \gamma _{ij}}\frac{\delta S_{%
\mathrm{b}}}{\delta \gamma _{kl}}\gamma _{ik}\gamma _{jl}-\frac{1}{d-1}\frac{%
\delta S_{\mathrm{a}}}{\delta \gamma _{ij}}\gamma _{ij}\frac{\delta S_{%
\mathrm{b}}}{\delta \gamma _{kl}}\gamma _{kl}+\frac{1}{4}G^{IJ}\frac{\delta
S_{\mathrm{a}}}{\delta \Phi ^{I}}\frac{\delta S_{\mathrm{b}}}{\delta \Phi
^{J}}\right) .
\end{equation}%
Keep in mind the finiteness of $S_{\mathrm{ren}}$ and the asymptotic
behavior of the fields\footnote{%
When $m_{I}^{2}=-\frac{d^{2}}{4}$, the leading behaviour of $\Phi ^{I}$ is
given by \cite{Bianchi0112,Martelli0205} $\Phi ^{I}\simeq \bar{\Phi}^{I}re^{-%
\frac{1}{2}dr}$ instead of eq. (\ref{jjxw}). Nevertheless, the remaining
derivation of the\ CPHJ equation is still valid.}%
\begin{equation}
\gamma _{ij}\simeq e^{2r}\bar{\gamma}_{ij},\;\qquad \Phi ^{I}\simeq
e^{-(d-\Delta _{I})r}\bar{\Phi}^{I},  \label{jjxw}
\end{equation}%
where $\bar{\gamma}_{ij}$ and $\bar{\Phi}^{I}$ are the sources on the field
theory and $\Delta _{I}=\frac{d}{2}+\sqrt{\frac{d^{2}}{4}+m_{I}^{2}}$ is the
conformal dimension. One can see immediately that $\{S_{\mathrm{ren}},S_{%
\mathrm{ren}}\}$\ vanishes as $r\rightarrow \infty $. Furthermore, at
leading order, we have%
\begin{eqnarray}
-\frac{\delta S_{\mathrm{ct}}}{\delta \gamma _{ij}} &\simeq &\frac{\partial L%
}{\partial \dot{\gamma}_{ij}}=\frac{1}{2\kappa ^{2}}\sqrt{\gamma }%
(K^{ij}-K\gamma ^{ij})\simeq -\frac{1}{2\kappa ^{2}}\sqrt{\gamma }%
(d-1)\gamma ^{ij},  \label{twomomenta1} \\
-\frac{\delta S_{\mathrm{ct}}}{\delta \Phi ^{I}} &\simeq &\frac{\partial L}{%
\partial \dot{\Phi}^{I}}=\frac{1}{\kappa ^{2}}\sqrt{\gamma }G_{IJ}\dot{\Phi}%
^{J}\simeq -\frac{1}{\kappa ^{2}}\sqrt{\gamma }G_{IJ}(d-\Delta _{I})\Phi
^{J}.  \label{twomomenta2}
\end{eqnarray}%
Substituting them into eq. (\ref{Hren}) gives\footnote{%
Here and below, we have considered that $S_{\mathrm{ren}}$ can be taken as
the functionals of ($\bar{\gamma}_{kl},\bar{\Phi}^{I}$) and ($\gamma
_{kl},\Phi ^{I},r$) from the viewpoints of the field theory and its gravity
dual, respectively.}%
\begin{eqnarray}
H_{\mathrm{ren}} &\simeq &\int_{\partial M}d^{d}x\left[ 2\frac{\delta S_{%
\mathrm{ren}}}{\delta \gamma _{ij}}\gamma _{ij}-(d-\Delta _{I})\Phi ^{I}%
\frac{\delta S_{\mathrm{ren}}}{\delta \Phi ^{I}}\right]   \nonumber \\
&\simeq &\int_{\partial M}d^{d}x\left[ 2\frac{\delta S_{\mathrm{ren}}}{%
\delta \bar{\gamma}_{kl}}\frac{\partial \left( e^{-2r}\gamma _{kl}\right) }{%
\partial \gamma _{ij}}\gamma _{ij}-(d-\Delta _{I})\Phi ^{I}\frac{\delta S_{%
\mathrm{ren}}}{\delta \bar{\Phi}^{J}}\frac{\partial \left( e^{(d-\Delta
_{I})r}\Phi ^{J}\right) }{\partial \Phi ^{I}}\right]   \nonumber \\
&\simeq &\int_{\partial M}d^{d}x\left[ 2\frac{\delta S_{\mathrm{ren}}}{%
\delta \bar{\gamma}_{kl}}\bar{\gamma}_{ij}-(d-\Delta _{I})\bar{\Phi}^{I}%
\frac{\delta S_{\mathrm{ren}}}{\delta \bar{\Phi}^{I}}\right] .  \label{Hren3}
\end{eqnarray}%
It exactly cancels%
\begin{eqnarray}
\frac{\partial S_{\mathrm{ren}}}{\partial r} &\simeq &\int_{\partial M}d^{d}x%
\left[ \frac{\delta S_{\mathrm{ren}}}{\delta \bar{\gamma}_{ij}}\frac{%
\partial \left( e^{-2r}\gamma _{ij}\right) }{\partial r}+\frac{\delta S_{%
\mathrm{ren}}}{\delta \bar{\Phi}^{I}}\frac{\partial \left( e^{(d-\Delta
_{I})r}\Phi ^{I}\right) }{\partial r}\right]   \nonumber \\
&\simeq &\int_{\partial M}d^{d}x\left[ -2\frac{\delta S_{\mathrm{ren}}}{%
\delta \bar{\gamma}_{ij}}\bar{\gamma}_{ij}+(d-\Delta _{I})\bar{\Phi}^{I}%
\frac{\delta S_{\mathrm{ren}}}{\delta \bar{\Phi}^{I}}\right] ,
\end{eqnarray}%
that is,%
\begin{equation}
H_{\mathrm{ren}}+\frac{\partial S_{\mathrm{ren}}}{\partial r}\simeq 0.
\label{RPHJ}
\end{equation}%
Thus, the complete HJ equation (\ref{HJ0}) has been reduced to the CPHJ
equation (\ref{HJdiv}).

To compare what we have done with previous references, some remarks are in
order. First, the separation of the on-shell action (\ref{S2}) is different
from dBVV's approach \cite{deBoer9912}. Our $-S_{\mathrm{ct}}$ includes all
the divergent terms but $S_{\mathrm{loc}}$ in eq. (14) of \cite{deBoer9912}
does not involve the logarithmic divergences. Second, in eq. (\ref%
{twomomenta1}) and eq. (\ref{twomomenta2}), we have used the well known
equality between two forms of canonical momenta at leading order. Equation (%
\ref{twomomenta2}) is nothing but the step 2 of the algorithm in \cite%
{Elvang1603}, which is taken as a shortcut to fix some coefficients of the
ansatz. Third, the CPHJ equation (\ref{HJdiv}) is not a completely new
result. In fact, a similar equation\footnote{%
To the best of our knowedge, the definition
(\ref{Hren}) of $H_{\text{\textrm{ren}}}$ is new and looks very different
from eq. (24) in \cite{Papadimitriou2016}. But in terms of eq. (\ref{jjxw})
and the first line of eq. (\ref{Hren3}), they are consistent indeed.} has
been given\ by eq. (27) in \cite{Papadimitriou2016} using the Hamiltonian
formulation of the renormalization group of local quantum field theories
\cite{Dolan9406}.
Also, eqs. (\ref{HJdiv}) and (\ref{RPHJ}) can be understood by the fact that
both $S_{\mathrm{ren}}$ and $S_{\mathrm{ct}}$ produce a canonical
transformation which can be associated with a Hamiltonian flow \cite{Papadimitriou1007}. Moreover, it should be stressed that our derivation is
similar to the part of the derivation of the dilatation operator method. In
particular, the first line of (\ref{Hren3}) equals to the dilatation
operator acting on $S_{\mathrm{ren}}$ and eq. (\ref{RPHJ}) can be related to
eq. (133) in \cite{Papadimitriou2016}. Our contribution here is to provide
a direct illustration of eq. (\ref{HJdiv}) by holography and point out that it can be taken as a
master equation to implement the holographic renormalization.

\section{Massive gravity}

We will study the massive gravity where the only dynamical field is the
spacetime metric and the boundary is supposed to be the AdS at infinity. We
will show that the CPHJ equation can be applied to the holographic
renormalization of massive gravity. Our target boundary dimensions are the
most interesting cases: $d=2,3,4$. The renormalization procedure for massive
gravity is only slightly different from the one for the Einstein-scalar
theory, which is given in Appendix A. We recommend reading it first since we
will neglect some similar details here.

\subsection{Hamilton-Jacobi formalism}

Consider the massive gravity with the action \cite{Vegh1301}
\begin{equation}
S_{MG}=-\frac{1}{2\kappa ^{2}}\int_{M}d^{d+1}x\sqrt{g}\Big(%
R[g]+d(d-1)+m^{2}\sum\limits_{n=1}^{4}\beta _{n}e_{n}(\mathcal{X})\Big)-%
\frac{1}{\kappa ^{2}}\int_{\partial M}d^{d}x\sqrt{\gamma }K.
\label{MG action}
\end{equation}%
The mass terms are constructed subtly to avoid the Boulware-Deser ghost,
where $\beta _{n}$ are constants and we will reparameterize them by $\alpha
_{n}=m^{2}\beta _{n}$. The characteristic tensor ${\mathcal{X}^{\mu }}_{\nu
} $ is defined as the square root of $g^{\mu \lambda }f_{\lambda \nu }$.
Here $g^{\mu \lambda }$ and $f_{\lambda \nu }$ are the dynamical and
reference metric, respectively. $e_{n}(\mathcal{X})$ are symmetric
polynomials of the eigenvalues of the $(d+1)\times (d+1)$ matrix ${\mathcal{X%
}^{\mu }}_{\nu }$:%
\begin{eqnarray}
e_{1} &=&[\mathcal{X}],\quad e_{2}=[\mathcal{X}]^{2}-[\mathcal{X}^{2}],\quad
e_{3}=[\mathcal{X}]^{3}-3[\mathcal{X}][\mathcal{X}^{2}]+2[\mathcal{X}^{3}],
\nonumber \\
e_{4} &=&[\mathcal{X}]^{4}-6[\mathcal{X}]^{2}[\mathcal{X}^{2}]+8[\mathcal{X}%
^{3}][\mathcal{X}]+3[\mathcal{X}^{2}]^{2}-6[\mathcal{X}^{4}],
\end{eqnarray}%
where we denote $[\mathcal{X}]={\mathcal{X}^{\mu }}_{\mu }$. The reference
metric can have various forms. Here we focus on
\begin{equation}
f_{\mu \nu }=\delta _{\mu }^{i}\delta _{\nu }^{j}f_{ij}  \label{fuv}
\end{equation}%
with $f_{ti}=0$, which is popular in the application of holography \cite%
{Vegh1301,Davison2013,Blake2013,Blake2014,Baggioli2014}.

When the Hamiltonian formulation is implemented in massive gravity, one may
encounter a complication. Massive gravity explicitly breaks the
diffeomorphism symmetry, which indicates that one cannot fix the gauge (\ref%
{gauge0}) in the whole bulk spacetime. These extra degrees of freedom,%
\footnote{%
We only need to worry about the effect of the shift vector. The lapse
function can be gauged away since the reference metric (\ref{fuv}) we choose
does not break the diffeomorphism symmetry along the radial direction.} if
involved, would complicate the gravitational Hamiltonian, the relevant constraint, and the sequent
holographic renormalization.
For the sake of simplicity, the GNC is assumed in the neighborhood of the boundary
and some additional conditions on ${\mathcal{X}^{\mu }}_{\nu }$ are imposed
in ref. \cite{Cao1509}. Here we release the conditions but still assume that
the GNC can be selected in a certain region near the boundary, that is,%
\begin{equation}
ds^{2}=dr^{2}+\gamma _{ij}dx^{i}dx^{j}.  \label{GNC}
\end{equation}%
More explicitly, we assume that the shift vector is falling off fast enough
asymptotically so that it does not affect the counterterms. This assumption
cannot be justified in general, but in Section 3.4, we will show some
interesting situations where it is true.

We would like to rewrite the mass terms by the boundary metric $\gamma _{ij}$%
. For this aim, let's define a tensor ${X^{i}}_{j}$ by%
\begin{equation}
{X^{i}}_{k}{X^{k}}_{j}=\gamma ^{ik}f_{kj}.
\end{equation}%
Due to eqs. (\ref{GNC}) and (\ref{fuv}), we have $[\mathcal{X}^{n}]=[X^{n}]$
and thereby $e_{n}(\mathcal{X})=e_{n}(X)$.

We proceed to study the CPHJ equation for massive gravity. Similar to the
derivation of eq. (\ref{HMS}) in Appendix A, one can obtain the Hamiltonian
for massive gravity by a Legendre transformation of the Lagrangian%
\begin{equation}
H\equiv \int_{\partial M}d^{d}x\pi ^{ij}\dot{\gamma}_{ij}-L=\int_{\partial
M}d^{d}x\mathcal{H},
\end{equation}%
where%
\begin{eqnarray}
L &=&-\frac{1}{2\kappa ^{2}}\int_{\partial M}d^{d}x\sqrt{\gamma }\left[
R+K^{2}-K_{ij}K^{ij}+d(d-1)+\sum\limits_{n=1}^{4}\alpha _{n}e_{n}(X)\right] ,
\\
\mathcal{H} &=&\frac{2\kappa ^{2}}{\sqrt{\gamma }}\left( \pi _{ij}\pi ^{ij}-%
\frac{1}{d-1}\pi ^{2}\right) +\frac{\sqrt{\gamma }}{2\kappa ^{2}}\left[
R+d(d-1)+\sum\limits_{n=1}^{4}\alpha _{n}e_{n}(X)\right] .  \label{HMG}
\end{eqnarray}%
Note that we have been working in the GNC. With the Hamiltonian in hands,
the CPHJ equation (\ref{HJdiv}) for massive gravity can be built up
following the same procedure in Section 2. Furthermore, it can be changed
into the form similar to eq. (\ref{sampleHJ}):%
\begin{equation}
R+\mathcal{K}+d(d-1)+\sum\limits_{n=1}^{4}\alpha _{n}e_{n}(X)+2\frac{%
\partial U}{\partial r}=0,  \label{MGHJ}
\end{equation}%
where%
\begin{equation}
\mathcal{K}=4Y_{ij}Y^{ij}-U^{2}-\frac{1}{d-1}(U-2Y)^{2},
\end{equation}%
and the definitions of $U$ and $Y_{ij}$ can be obtained from eqs. (\ref{U})
and (\ref{Yij}) with vanishing scalar fields.

\subsection{Action ansatz and variation}

The main difference that we mentioned at the beginning of this section
resides in the inverse metric expansion of $U$.\footnote{%
Usually, the derivative expansion is equivalent to the inverse metric
expansion. But for massive gravity, they are different and the latter is
more convenient.} The definition of ${X^{i}}_{j}$ suggests that the
counterterms in massive gravity may contain the terms with half-integer
inverse metrics, that is,
\begin{equation}
U=U_{(0)}+U_{(1)}+\cdots +U_{(d)},\qquad d=2,3,4,
\end{equation}%
where $U_{(2k)}$ contains $k$ inverse metrics. The sufficient ansatz for
each order is%
\begin{eqnarray}
U_{(0)} &=&A(r),  \nonumber \\
U_{(1)} &=&B(r)[X],  \nonumber \\
U_{(2)} &=&C_{1}(r)R+C_{2}(r)[X^{2}]+C_{3}(r)[X]^{2},  \nonumber \\
U_{(3)}
&=&D_{1}(r)[X]R+D_{2}(r)R_{ij}X^{ij}+D_{3}(r)[X^{3}]+D_{4}(r)[X^{2}][X]+D_{5}(r)[X]^{3},
\nonumber \\
U_{(4)}
&=&E_{1}(r)R_{ij}R^{ij}+E_{2}(r)R^{2}+E_{3}(r)[X^{2}]R+E_{4}(r)[X]^{2}R+E_{5}(r)R_{ij}X^{ij}[X]
\nonumber \\
&&+E_{6}(r){R^{i}}_{j}{X^{j}}_{k}{X^{k}}%
_{i}+E_{7}(r)[X^{4}]+E_{8}(r)[X^{3}][X]+E_{9}(r)[X^{2}][X]^{2}  \nonumber \\
&&+E_{10}(r)[X^{2}]^{2}+E_{11}(r)[X]^{4}+E_{12}(r)X^{ij}\nabla ^{k}\nabla
_{j}X_{ki}+E_{13}(r)X^{ij}\nabla ^{k}\nabla _{k}X_{ij}  \nonumber \\
&&+E_{14}(r)[X]\nabla _{i}\nabla _{j}X^{ij}+E_{15}(r)[X]\nabla _{i}\nabla
^{i}[X]+\cdots ,  \label{MGansatz}
\end{eqnarray}%
where \textquotedblleft $\cdots $\textquotedblright\ denote the terms which
can be related to the existed terms by total derivatives (like the term $%
\sim \nabla _{i}X^{ij}\nabla _{j}[X]$) or which turns out to have the
vanishing coefficients finally (like the term $\sim X^{ij}\nabla _{j}\nabla
^{k}X_{ki}$). We will explain this issue later.

Taking the variation of the action ansatz with respect to the boundary
metric, we can obtain each term in the expansion of $\mathcal{K}$%
\begin{equation}
\mathcal{K}\mathcal{=}\mathcal{K}_{(0)}+\mathcal{K}_{(1)}+\cdots +\mathcal{K}%
_{(d)}.
\end{equation}%
The detail of computation is presented in Appendix B. Here we write down the
results%
\begin{eqnarray}
\mathcal{K}_{(0)} &=&-\frac{d}{d-1}U_{(0)}^{2}, \\
\mathcal{K}_{(1)} &=&-2U_{(0)}U_{(1)}-\frac{2}{d-1}U_{(0)}\left(
U_{(1)}-2Y_{(1)}\right) ,  \nonumber \\
\mathcal{K}_{(2)} &=&4Y_{(1)ij}Y_{(1)}^{ij}-\left(
2U_{(0)}U_{(2)}+U_{(1)}^{2}\right) -\frac{1}{d-1}\left(
2U_{(0)}(U_{(2)}-2Y_{(2)})+(U_{(1)}-2Y_{(1)})^{2}\right) ,  \nonumber \\
\mathcal{K}_{(3)} &=&4\left(
Y_{(1)ij}Y_{(2)}^{ij}+Y_{(2)ij}Y_{(1)}^{ij}\right) -2\left(
U_{(0)}U_{(3)}+U_{(1)}U_{(2)}\right)  \nonumber \\
&&-\frac{2}{d-1}\Big(%
U_{(0)}(U_{(3)}-2Y_{(3)})+(U_{(1)}-2Y_{(1)})(U_{(2)}-2Y_{(2)})\Big),
\nonumber \\
\mathcal{K}_{(4)} &=&4\left(
Y_{(1)ij}Y_{(3)}^{ij}+Y_{(1)}^{ij}Y_{(3)ij}+Y_{(2)ij}Y_{(2)}^{ij}\right)
-\left( 2U_{(0)}U_{(4)}+2U_{(1)}U_{(3)}+U_{(2)}^{2}\right)  \nonumber \\
&&-\frac{1}{d-1}\Big(%
2U_{(0)}(U_{(4)}-2Y_{(4)})+2(U_{(1)}-2Y_{(1)})(U_{(3)}-2Y_{(3)})+(U_{(2)}-2Y_{(2)})^{2}%
\Big),  \nonumber
\end{eqnarray}%
where the expressions of $Y_{(m)ij}Y_{(n)}^{ij}$ can be readily obtained
from eq. (\ref{YY}) and $Y_{(k)}$ can be related to $U_{(k)}$ by eqs. (\ref%
{Y1U1}), (\ref{Y2U2}), (\ref{Y3U3}) and (\ref{Y4U4}).

\subsection{Solution of CPHJ equation}

We proceed to solve the CPHJ equation (\ref{MGHJ}) iteratively to determine
the unknown coefficients $\left( A,B,C_{i},D_{i},E_{i}\right) $.

\begin{itemize}
\item The order 0 descent equation is
\begin{equation}
\mathcal{K}_{(0)}+d(d-1)+2\frac{\partial U_{(0)}}{\partial r}=0,
\end{equation}
which has the solution
\begin{equation}
A(r)=-(d-1)+\mathcal{O}(e^{-dr}).
\end{equation}
We only keep the leading term. By power counting, one can see that the
subleading term is not divergent.

\item With the order 0 result, one is able to solve the order 1 descent
equation
\begin{equation}
{\mathcal{K}}_{(1)}+\alpha _{1}e_{1}+2\frac{\partial U_{(1)}}{\partial r}=0.
\label{kappa1}
\end{equation}%
The solution about $[X]$ is
\begin{equation}
B(r)=\frac{\alpha _{1}}{2(1-d)}+\mathcal{O}(e^{(1-d)r}).
\end{equation}%
So $U_{(2k)}$ does contain the term with half-integer $k$.

\item It is turned to deal with the order 2 descent equation
\begin{equation}
{R+\mathcal{K}}_{(2)}+sgn(d-2)\alpha _{2}e_{2}+2\frac{\partial U_{(2)}}{%
\partial r}=0,
\end{equation}%
which is needed when $d\geq 2$. Here we have introduced the sign function
\begin{equation}
sgn(x)=\begin{cases} 0 & x=0 \\ 1 & x>0\end{cases},
\end{equation}%
which is invoked to emphasize the polynomial $e_{d}(X)=0$ under the choice $%
f_{t\mu }=0$, as presented in (\ref{edx}). The independence of boundary
conditions results in
\begin{eqnarray}
R &:&\quad 1+2(d-2)C_{1}+2\dot{C}_{1}=0,  \nonumber \\
\lbrack X^{2}] &:&\quad \bar{B}^{2}+2(d-2)C_{2}-sgn(d-2)\alpha _{2}+2\dot{C}%
_{2}=0,  \nonumber \\
\lbrack X]^{2} &:&\quad 2(d-2)C_{3}-\bar{B}^{2}+sgn(d-2)\alpha _{2}+2\dot{C}%
_{3}=0.
\end{eqnarray}%
Here $\bar{B}$ is defined as a constant, denoting the solved but unfixed
coefficient $B$. Later notations about $\bar{C}_{i}$ and $\bar{D}_{i}$ are
similar. The above equations have the solutions:
\begin{eqnarray}
C_{1} &\!\!=\!\!&\begin{cases} \frac{1}{2(2-d)}+\mathcal{O}(e^{(2-d)r}),\; &
d>2 \\ -\frac{r}{2}+\mathcal{O}(1), & d=2\end{cases},\qquad C_{2}=%
\begin{cases} \frac{\bar{B}^{2}-\alpha
_{2}}{2(2-d)}+\mathcal{O}(e^{(2-d)r}),\; & d>2 \\
-\frac{\bar{B}^{2}}{2}r+\mathcal{O}(1), & d=2\end{cases}  \nonumber \\
C_{3} &\!\!=\!\!&\begin{cases} -\frac{\bar{B}^{2}-\alpha
_{2}}{2(2-d)}+\mathcal{O}(e^{(2-d)r}),\; & d>2 \\
\frac{\bar{B}^{2}}{2}r+\mathcal{O}(1), & d=2\end{cases}
\end{eqnarray}

\item So far we have determined all the divergent terms for $d=2$ but not
enough for $d=3,4$. The next is the order 3 descent equation
\begin{equation}
{\mathcal{K}}_{(3)}+sgn(d-3)\alpha _{3}e_{3}+2\frac{\partial U_{(3)}}{%
\partial r}=0.
\end{equation}%
Collecting various functional terms gives
\begin{eqnarray}
\lbrack X]R &:&\quad 2(d-3)D_{1}-2\bar{B}\bar{C}_{1}+2\dot{D}_{1}=0,
\nonumber \\
R_{ij}X^{ij} &:&\quad 4\bar{B}\bar{C}_{1}+2(d-3)D_{2}+2\dot{D}_{2}=0,
\nonumber \\
\lbrack X^{3}] &:&\quad 4\bar{B}\bar{C}_{2}+2(d-3)D_{3}+sgn(d-3)2\alpha
_{3}+2\dot{D}_{3}=0,  \nonumber \\
\lbrack X^{2}][X] &:&\quad 2(d-3)D_{4}-6\bar{B}\bar{C}_{2}-sgn(d-3)3\alpha
_{3}+2\dot{D}_{4}=0,  \nonumber \\
\lbrack X]^{3} &:&\quad 2(d-3)D_{5}+2\bar{B}\bar{C}_{2}+sgn(d-3)\alpha _{3}+2%
\dot{D}_{5}=0.
\end{eqnarray}%
where we have used ${C}_{3}=-{C}_{2}$. The solutions are
\begin{eqnarray}
D_{1} &\!\!\!=\!\!\!&\begin{cases}
\frac{\bar{B}\bar{C}_{1}}{d-3}+\mathcal{O}(e^{(3-d)r}),\; & d>3 \\
\bar{B}\bar{C}_{1}r+\mathcal{O}(1), & d=3\end{cases},\quad \qquad D_{2}=%
\begin{cases} -2\frac{\bar{B}\bar{C}_{1}}{d-3}+\mathcal{O}(e^{(3-d)r}),\quad
& d>3 \\ -2\bar{B}\bar{C}_{1}r+\mathcal{O}(1), & \ d=3\end{cases}  \nonumber
\\
D_{3} &\!\!\!=\!\!\!&\begin{cases} \frac{2\bar{B}\bar{C}_{2}+\alpha
_{3}}{3-d}+\mathcal{O}(e^{(3-d)r}),\; & d>3 \\
-2\bar{B}\bar{C}_{2}r+\mathcal{O}(1), & d=3\end{cases},\quad D_{4}=%
\begin{cases} -\frac{3}{2}\frac{2\bar{B}\bar{C}_{2}+\alpha
_{3}}{3-d}+\mathcal{O} (e^{(3-d)r}),\; & d>3 \\
3\bar{B}\bar{C}_{2}r+\mathcal{O}(1), & d=3\end{cases}  \nonumber \\
D_{5} &\!\!\!=\!\!\!&\begin{cases}
\frac{1}{2}\frac{2\bar{B}\bar{C}_{2}+\alpha _{3}}{3-d}+\mathcal{O}
(e^{(3-d)r}),\; & d>3 \\ -\bar{B}\bar{C}_{2}r+\mathcal{O}(1), &
d=3\end{cases}.
\end{eqnarray}%
Specifically, one can read ${D}_{2}=-2{D}_{1}$, ${D}_{3}=2{D}_{5}$, ${D}%
_{4}=-3{D}_{5}$.

\item Now the case $d=3$ is completed. Let us deal with the order 4 descent
equation
\begin{equation}
{\mathcal{K}}_{(4)}+sgn(d-4)\alpha _{4}e_{4}+2\frac{\partial U_{(4)}}{%
\partial r}=0.  \label{MG4HJ}
\end{equation}%
It induces a series of equations%
\begin{eqnarray}
R_{ij}R^{ij} &:&\quad 4\bar{C}_{1}^{2}+2(d-4)E_{1}+2\dot{E}_{1}=0,  \nonumber
\\
R^{2} &:&\quad -\frac{d}{d-1}\bar{C}_{1}^{2}+2(d-4)E_{2}+2\dot{E}_{2}=0,
\end{eqnarray}%
\begin{eqnarray}
\lbrack X^{2}]R &:&\quad 2\bar{B}\bar{D}_{1}-\frac{d}{d-1}2\bar{C}_{1}\bar{C}%
_{2}+2(d-4)E_{3}+2\dot{E}_{3}=0,  \nonumber \\
\lbrack X]^{2}R &:&\quad -2\bar{B}\bar{D}_{1}+\frac{d}{d-1}2\bar{C}_{1}\bar{C%
}_{2}+2(d-4)E_{4}+2\dot{E}_{4}=0,  \nonumber \\
R_{ij}X^{ij}[X] &:&\quad 8\bar{B}\bar{D}_{1}-8\bar{C}_{1}\bar{C}%
_{2}+2(d-4)E_{5}+2\dot{E}_{5}=0,  \nonumber \\
{R^{i}}_{j}{X^{j}}_{k}{X^{k}}_{i} &:&\quad -12\bar{B}\bar{D}_{1}+8\bar{C}_{1}%
\bar{C}_{2}+2(d-4)E_{6}+2\dot{E}_{6}=0,
\end{eqnarray}%
\begin{eqnarray}
\lbrack X^{4}] &:&\quad 12\bar{B}\bar{D}_{5}+4\bar{C}_{2}^{2}-sgn(d-4)6%
\alpha _{4}+2(d-4)E_{7}+2\dot{E}_{7}=0,  \nonumber \\
\lbrack X^{3}][X] &:&\quad -16\bar{B}\bar{D}_{5}-8\bar{C}_{2}^{2}+sgn(d-4)8%
\alpha _{4}+2(d-4)E_{8}+2\dot{E}_{8}=0,  \nonumber \\
\lbrack X^{2}][X]^{2} &:&\quad 12\bar{B}\bar{D}_{5}+\left( 4+2\frac{d}{d-1}%
\right) \bar{C}_{2}^{2}-sgn(d-4)6\alpha _{4}+2(d-4)E_{9}+2\dot{E}_{9}=0,
\nonumber \\
\lbrack X^{2}]^{2} &:&\quad -6\bar{B}\bar{D}_{5}-\frac{d}{d-1}\bar{C}%
_{2}^{2}+sgn(d-4)3\alpha _{4}+2(d-4)E_{10}+2\dot{E}_{10}=0,  \nonumber \\
\lbrack X]^{4} &:&\quad -2\bar{B}\bar{D}_{5}-\frac{d}{d-1}\bar{C}%
_{2}^{2}+sgn(d-4)\alpha _{4}+2(d-4)E_{11}+2\dot{E}_{11}=0,
\end{eqnarray}%
\begin{eqnarray}
X^{ij}\nabla ^{k}\nabla _{j}X_{ki} &:&\quad 8\bar{B}\bar{D}%
_{1}+2(d-4)E_{12}+2\dot{E}_{12}=0,  \nonumber \\
X^{ij}\nabla ^{k}\nabla _{k}X_{ij} &:&\quad -4\bar{B}\bar{D}%
_{1}+2(d-4)E_{13}+2\dot{E}_{13}=0,  \nonumber \\
\lbrack X]\nabla _{i}\nabla _{j}X^{ij} &:&\quad -8\bar{B}\bar{D}%
_{1}+2(d-4)E_{14}+2\dot{E}_{14}=0,  \nonumber \\
\lbrack X]\nabla _{i}\nabla ^{i}[X] &:&\quad 4\bar{B}\bar{D}%
_{1}+2(d-4)E_{15}+2\dot{E}_{15}=0.
\end{eqnarray}%
Although there are so many equations, their solutions are still simple. When
$d=4$, they are
\begin{eqnarray}
E_{1} &=&-2\bar{C}_{1}^{2}r+\mathcal{O}(1),\qquad E_{2}=\frac{2}{3}\bar{C}%
_{1}^{2}r+\mathcal{O}(1),  \nonumber \\
E_{3} &=&\left( \frac{4}{3}\bar{C}_{1}\bar{C}_{2}-\bar{B}\bar{D}_{1}\right)
r+\mathcal{O}(1),\qquad E_{4}=\left( -\frac{4}{3}\bar{C}_{1}\bar{C}_{2}+\bar{%
B}\bar{D}_{1}\right) r+\mathcal{O}(1),  \nonumber \\
E_{5} &=&(4\bar{C}_{1}\bar{C}_{2}-4\bar{B}\bar{D}_{1})r+\mathcal{O}%
(1),\qquad E_{6}=(-4\bar{C}_{1}\bar{C}_{2}+6\bar{B}\bar{D}_{1})r+\mathcal{O}%
(1),  \nonumber \\
E_{7} &=&(-2\bar{C}_{2}^{2}-6\bar{B}\bar{D}_{5})r+\mathcal{O}(1),\qquad
E_{8}=(4\bar{C}_{2}^{2}+8\bar{B}\bar{D}_{5})r+\mathcal{O}(1),  \nonumber \\
E_{9} &=&\left( -\frac{4}{3}\bar{C}_{2}^{2}-2\bar{C}_{2}^{2}-6\bar{B}\bar{D}%
_{5}\right) r+\mathcal{O}(1),\qquad E_{10}=\left( \frac{2}{3}\bar{C}%
_{2}^{2}+3\bar{B}\bar{D}_{5}\right) r+\mathcal{O}(1),  \nonumber \\
E_{11} &=&\left( \frac{2}{3}\bar{C}_{2}^{2}+\bar{B}\bar{D}_{5}\right) r+%
\mathcal{O}(1),\qquad E_{12}=-4\bar{B}\bar{D}_{1}r+\mathcal{O}(1),  \nonumber
\\
E_{13} &=&2\bar{B}\bar{D}_{1}r+\mathcal{O}(1),\qquad E_{14}=4\bar{B}\bar{D}%
_{1}r+\mathcal{O}(1),  \nonumber \\
E_{15} &=&-2\bar{B}\bar{D}_{1}r+\mathcal{O}(1).
\end{eqnarray}%
Particularly we notice the simplification
\begin{eqnarray}
&&E_{7}[X^{4}]+E_{8}[X^{3}][X]+E_{9}[X^{2}][X]^{2}+E_{10}[X^{2}]^{2}+E_{11}[X]^{4}
\nonumber \\
&=&\bigg\{\bar{B}\bar{D}%
_{5}(-6[X^{4}]+8[X^{3}][X]-6[X^{2}][X]^{2}+3[X^{2}]^{2}+[X]^{4})  \nonumber
\\
&&+\frac{2}{3}\bar{C}_{2}^{2}([X^{2}]^{2}-2[X^{2}][X]^{2}+[X]^{4})-\frac{1}{3%
}\bar{C}_{2}^{2}(6[X^{4}]-12[X^{3}][X]+6[X^{2}][X]^{2})\bigg\}r  \nonumber \\
&=&\bigg\{\bar{B}\bar{D}_{5}e_{4}+\frac{2}{3}\bar{C}_{2}^{2}e_{2}^{2}-\frac{1%
}{3}\bar{C}_{2}^{2}(-2e_{1}e_{3}+3e_{2}^{2}-e_{4})\bigg\}r  \nonumber \\
&=&\left( \frac{2}{3}e_{1}e_{3}-\frac{1}{3}e_{2}^{2}\right) \bar{C}_{2}^{2}r.
\end{eqnarray}
\end{itemize}

Obviously, the number of divergent terms increases quickly when the
spacetime dimension increases. Here we give a remark that is useful to avoid
neglecting certain divergent terms. Suppose that there should be a real
divergent term labeled by $a(r)\mathcal{A}_{(k)}$ in the ansatz $U_{(k)}$
and $\mathcal{A}_{(k)}$ does not appear in $\mathcal{B}$, where $\mathcal{B}$
contains every term in $H_{\mathrm{ct}}$ except $\mathcal{K}$. Then we write
the order $k$ CPHJ equation as:%
\begin{equation}
\mathcal{K}_{(k)}+\mathcal{B}_{(k)}+2\frac{\partial U_{(k)}}{\partial r}=0,
\label{kth}
\end{equation}%
where%
\begin{equation}
\mathcal{K}_{(k)}=\sum\limits_{\substack{ m+n=k,  \\ 0<m,n<k}}%
4Y_{(m)ij}Y_{(n)}^{ij}-\sum\limits_{m+n=k}U_{(m)}U_{(n)}-\frac{1}{d-1}%
\sum\limits_{m+n=k}(U_{(m)}-2Y_{(m)})(U_{(n)}-2Y_{(n)}).
\end{equation}%
We proceed to present an assumption that will be falsified in the end: $%
\mathcal{A}_{(k)}$ only appears in $U_{(k)}$ in eq. (\ref{kth}). Setting $m\
(\text{or }n)=0$ and using $Y_{(k)}=\frac{k}{2}U_{(k)}+\text{total
derivatives}$, we have
\begin{equation}
\mathcal{A}_{(k)}:\quad (d-k)a+\dot{a}=0.
\end{equation}%
The solution is $a=\mathcal{O}(e^{(k-d)r})$, which is impossible for a real
divergent term because: the scaling of $\sqrt{\gamma }a(r)\mathcal{A}_{(k)}$
is $e^{dr}\cdot e^{(k-d)r}\cdot e^{-kr}=\mathcal{O}(1)$. Thus, our previous
assumption is invalid, that is, the terms other than $U_{(k)}$ and $\mathcal{%
\ B}_{(k)}$ in eq. (\ref{kth}) must contain $\mathcal{A}_{(k)}$ whose
coefficient is nonvanishing. Note that all these terms can be worked out
with the pre-solved $U_{(m)}$, where $m<k$.

Put it another way, suppose that one has accidentally neglected a real
divergent term $a(r)\mathcal{A}_{(k)}$ in the ansatz $U_{(k)}$. When
organizing the $k$th order CPHJ equation, one then will obtain an
ill-defined algebraic equation about the potentially divergent term $%
\mathcal{A}_{(k)}$. This is implied by the above analysis. Thus, the CPHJ
equation can remind one to add $a(r)\mathcal{A}_{(k)}$ which makes the
ansatz sufficient.

Keeping this remark in mind, we can explain quickly why the term like $%
E_{16}X^{ij}\nabla _{j}\nabla ^{k}X_{ki}$ in the ansatz $U_{(4)}$ is not
necessary. This is because in the 4th order CPHJ equation, $X^{ij}\nabla
_{j}\nabla ^{k}X_{ki}$ only appears in $U_{(4)}$.

Finally, we turn back to present the counterterm action by collecting above
results. It can be written as
\begin{equation}
S_{\mathrm{ct}}=-\frac{1}{\kappa ^{2}}\int_{\Sigma }d^{d}x\sqrt{\gamma }\big(%
U_{(0)}+U_{(1)}+...+U_{(d)}\big).
\end{equation}%
The first two terms have the uniform%
\begin{equation}
U_{(0)}=-(d-1),\qquad U_{(1)}=\frac{\alpha _{1}}{2(1-d)}e_{1}.
\end{equation}%
But other terms depend on the dimensions, which will be listed as follows.

\begin{itemize}
\item $d=2$
\begin{equation}
U_{(2)}=\left( -\frac{1}{2}R+\frac{\alpha _{1}^{2}}{8}([X]^{2}-[X^{2}])%
\right) r=-\frac{1}{2}Rr,
\end{equation}%
Here we have used $e_{d}=0$. Then the counterterm action is
\begin{equation}
S_{\mathrm{ct}}=\frac{1}{\kappa ^{2}}\int_{\Sigma }d^{2}x\sqrt{\gamma }%
\left( 1+\frac{1}{2}\alpha _{1}e_{1}+\frac{1}{2}Rr\right) .  \label{sctd2}
\end{equation}

\item $d=3$
\begin{eqnarray}
U_{(2)} &=&-\frac{1}{2}R+\left(\frac{1}{2}\alpha _{2}-\frac{1}{32}\alpha
_{1}^{2}\right)\left([X^{2}]-[X]^{2}\right)=-\frac{1}{2}R-\left(\frac{1}{2}%
\alpha _{2}-\frac{1}{32} \alpha _{1}^{2}\right)e_{2},  \nonumber \\
U_{(3)} &=&\left\{\frac{\alpha _{1}}{8}([X]R-2R_{ij}X^{ij})+\left(\frac{1}{8}%
\alpha _{1}\alpha _{2}-\frac{1}{128}\alpha
_{1}^{3}\right)(2[X^{3}]-3[X^{2}][X]+[X]^{3}) \right\}r  \nonumber \\
&=&\frac{\alpha _{1}}{8}(e_{1}R-2R_{ij}X^{ij})r.
\end{eqnarray}
\end{itemize}

The counterterm action is
\begin{equation}
S_{\mathrm{ct}}=\frac{1}{\kappa ^{2}}\int_{\Sigma }d^{3}x\sqrt{\gamma }%
\left\{2+\frac{1}{4}\alpha _{1}e_{1}+\frac{1}{2}R+\left(\frac{1}{2}\alpha
_{2}-\frac{1}{ 32}\alpha _{1}^{2}\right)e_{2}+\frac{\alpha _{1}}{8}%
(2R_{ij}X^{ij}-e_{1}R)r\right\}.  \label{SCT3}
\end{equation}%
This result is the same as eq. (3.15) in \cite{Cao1509} up to the last
logarithmic terms. Note that the logarithmic terms vanish precisely if one
takes the metric (B7) in \cite{Cao1509}.

\begin{itemize}
\item $d=4$
\begin{eqnarray}
U_{(2)} &=&-\frac{1}{4}R+\left(\frac{1}{4}\alpha _{2}-\frac{1}{144}\alpha
_{1}^{2}\right)([X^{2}]-[X]^{2})=-\frac{1}{4}R-\left(\frac{1}{4}\alpha _{2}-%
\frac{1}{144 }\alpha _{1}^{2}\right)e_{2}, \\
U_{(3)} &=&\frac{\alpha _{1}}{24}([X]R-2R_{ij}X^{ij})+\left(\frac{1}{24}%
\alpha _{1}\alpha _{2}-\frac{1}{864}\alpha _{1}^{3}-\frac{\alpha _{3}}{2}
\right)(2[X^{3}]-3[X^{2}][X]+[X]^{3})  \nonumber \\
&=&\frac{\alpha _{1}}{24}(e_{1}R-2R_{ij}X^{ij})+\left(\frac{1}{24}\alpha
_{1}\alpha _{2}-\frac{1}{864}\alpha _{1}^{3}-\frac{\alpha _{3}}{2}%
\right)e_{3}, \\
U_{(4)} &=&\bigg\{-\frac{1}{8}\left(R_{ij}R^{ij}-\frac{1}{3}%
R^{2}\right)-\left(\frac{\alpha _{1}^{2}}{108}-\frac{\alpha _{2}}{12}%
\right)e_{2}R+\left(\frac{5}{144}\alpha _{1}^{2}- \frac{\alpha _{2}}{4}%
\right)e_{1}R_{ij}X^{ij} \\
&&+\left(\frac{\alpha _{2}}{4}-\frac{7\alpha _{1}^{2}}{144}\right){R^{i}}_{j}%
{X^{j}}_{k} {X^{k}}_{i}-\left(\frac{1}{3}e_{2}^{2}-\frac{2}{3}%
e_{1}e_{3}\right)\left(\frac{\alpha _{2}}{4 }-\frac{\alpha _{1}^{2}}{144}%
\right)^{2}  \nonumber \\
&&-\frac{\alpha _{1}^{2}}{72}(-2X^{ij}\nabla ^{k}\nabla
_{j}X_{ki}+X^{ij}\nabla ^{k}\nabla _{k}X_{ij}+2[X]\nabla _{i}\nabla
_{j}X^{ij}-[X]\nabla _{i}\nabla ^{i}[X])\bigg\}r.  \nonumber
\end{eqnarray}
The counterterm action is
\begin{eqnarray}
S_{\mathrm{ct}} &=&\frac{1}{\kappa ^{2}}\int_{\Sigma }d^{4}x\sqrt{\gamma } %
\bigg\{3+\frac{1}{6}\alpha _{1}e_{1}+\frac{1}{4}R+\left(\frac{1}{4}\alpha
_{2}- \frac{1}{144}\alpha _{1}^{2}\right)e_{2}  \label{Sct4} \\
&&+\frac{1}{24}\alpha _{1}(2R_{ij}X^{ij}-e_{1}R)-\left(\frac{1}{24}\alpha
_{1}\alpha _{2}-\frac{1}{864}\alpha _{1}^{3}-\frac{\alpha _{3}}{2}%
\right)e_{3}  \nonumber \\
&&+\bigg[\frac{1}{8}\left(R_{ij}R^{ij}-\frac{1}{3}R^{2}\right)+\left(\frac{%
\alpha _{1}^{2}}{ 108}-\frac{\alpha _{2}}{12}\right)e_{2}R-\left(\frac{5}{144%
}\alpha _{1}^{2}-\frac{ \alpha _{2}}{4}\right)e_{1}R_{ij}X^{ij}  \nonumber \\
&&-\left(\frac{\alpha _{2}}{4}-\frac{7\alpha _{1}^{2}}{144}\right){R^{i}}_{j}%
{X^{j}}_{k} {X^{k}}_{i}+\left(\frac{1}{3}e_{2}^{2}-\frac{2}{3}%
e_{1}e_{3}\right)\left(\frac{\alpha _{2}}{4 }-\frac{\alpha _{1}^{2}}{144}%
\right)^{2}  \nonumber \\
&&+\frac{\alpha _{1}^{2}}{72}(-2X^{ij}\nabla ^{k}\nabla
_{j}X_{ki}+X^{ij}\nabla ^{k}\nabla _{k}X_{ij}+2[X]\nabla _{i}\nabla
_{j}X^{ij}-[X]\nabla _{a}\nabla ^{a}[X])\bigg]r\bigg\}.  \nonumber
\end{eqnarray}
\end{itemize}

\subsection{Renormalized action}

Now we will show that in some situations the divergent part of the on-shell
action is actually cancelled by the counterterms that we have derived. In
these situations, our assumption of the shift vector is justified.

\subsubsection{Background level}

Consider the background level at first.
Select the reference metric as%
\begin{equation}
f_{ij}=\mathrm{diag}(0,h_{ab}),
\end{equation}%
where $h_{ab}$ is the metric of a $\left( d-1\right) $-dimensional Einstein
space with constant curvature $\left( d-2\right) (d-1)k$ and the parameter $k=0,\pm 1$.
There are the black-hole solutions for massive gravity in $\left( d+1\right)
$-dimensional spacetimes
\begin{equation}
ds^{2}=f(z)d\tau ^{2}+f^{-1}(z)dz^{2}+z^{2}h_{ab}dx^{a}dx^{b},  \label{BHmetric0}
\end{equation}%
where the coordinate $z$ is related to $r$ via $z=e^{r}$ and the blackening
factor is%
\begin{equation}
f(z)=k+z^{2}-\frac{m_{0}}{z^{d-2}}+\frac{\alpha _{1}}{d-1}z+\alpha _{2}+%
\frac{(d-2)\alpha _{3}}{z},  \label{BHmetric}
\end{equation}%
with the mass parameter $m_{0}$.
There are no cross terms in eq. (\ref{BHmetric0}), so the GNC is obviously available. In the following, we calculate
the counterterms and the renormalized action $S_{\mathrm{ren}}=\lim\limits_{z\rightarrow \infty }(S_{\mathrm{on-shell}}+S_{\mathrm{ct}})$
for different dimensions.

\begin{itemize}
\item $d=2$

Using the background metric (\ref{BHmetric0}), the counterterms (\ref{sctd2})
can be reduced to

\begin{equation}
S_{\mathrm{ct}}=\frac{V}{2\kappa ^{2}}\left[
-(a_{1}+2z)\sqrt{f}+2f+zf^{\prime }\right] ,
\end{equation}%
where $V\equiv \int_{\Sigma }d^{d}x\sqrt{h}$ and $h$ is the determinant of
the metric $h_{ab}$. Then the renormalized action can be obtained%
\begin{equation}
S_{\mathrm{ren}}=\frac{V}{2\kappa
^{2}}\left( z_{+}^{2}-\frac{a_{1}}{4}\right) ,
\end{equation}%
where $z_{+}$ denotes the location of the
horizon.

\item $d=3$

When $d=3$, the finiteness of the renormalized action has been checked in
the black-hole background \cite{Cao1509}.

\item $d=4$

The counterterms (\ref{Sct4}) can be calculated as%
\begin{eqnarray}
S_{\mathrm{ct}} &=&-\frac{V}{2\kappa ^{2}}\bigg\{\left(
6z^{2}f+z^{3}f^{\prime }\right) -6z^{3}\sqrt{f}-\alpha _{1}z^{2}\sqrt{f}%
+\left( \frac{1}{12}\alpha _{1}^{2}z-3\alpha _{2}z-3kz\right) \sqrt{f}
\nonumber \\
&&+\left( \frac{1}{2}\alpha _{1}\alpha _{2}-6\alpha _{3}+\frac{1}{2}\alpha
_{1}k-\frac{1}{72}\alpha _{1}^{3}\right) \sqrt{f}\bigg\}.
\end{eqnarray}%
Appending the counterterms to the on-shell action, we have%
\begin{eqnarray}
S_{\mathrm{ren}} &=&-\frac{V}{2\kappa ^{2}}\Big\{\left(
z_{+}^{4}-kz_{+}^{2}-\alpha _{2}z_{+}^{2}-4\alpha _{3}z_{+}\right)  \nonumber
\\
&&-\frac{3}{4}k^{2}-\frac{3}{2}k\alpha _{2}-\frac{3}{4}\alpha
_{2}^{2}-\alpha _{1}\alpha _{3}+\frac{\alpha _{1}^{2}}{8}(\alpha _{2}+k)-%
\frac{5}{1728}\alpha _{1}^{4}\Big\}.
\end{eqnarray}
\end{itemize}

As shown, for various dimensions, the divergent terms in the on-shell action
at the background level have been canceled out. Moreover, we find that the
Hawking temperature $T=f^{\prime }(z_{+})/\left( 4\pi \right) $, the
Bekenstein entropy $S=4\pi z_{+}^{d-1}V/\left( 2\kappa ^{2}\right) $, and
the grand potential $\Omega =-TS_{\mathrm{ren}}$ exactly obey the
thermodynamical formula $\partial \Omega /\partial T=S$. This is a
self-consistent check of our results.

\subsubsection{Perturbation level}

At the perturbation level, we cannot prove in general that the shift vector is
falling off fast enough. Fortunately, for the optical perturbations (finite
frequency, zero wave vector) that are often studied in the holographic
theories of condensed matther physics, we find that our counterterms are
enough to cancel the divergence terms in some cases. To exhibit them
clearly, we turn on the time-dependent linear perturbations above the black-hole background (\ref{BHmetric0}). We focus on $%
k=0$ for simplicity, which denotes the flat geometry of the field theory.
These perturbation modes can be separated into three groups. The shift
vector appears as a vector mode but decouples with the scalar and tensor
modes. Thus, our counterterms are applicable for the theories involving the
scalar and tensor modes. As for the vector modes, we will show that the
shift vector is actually falling fast enough in three cases below. For
convenience, we write the coupled vector modes as $\delta g_{tx}\left(
t,z\right) =z^{2}h_{tx}(t,z)$ and $\delta g_{xz}\left( t,z\right)
=z^{2}h_{xz}(t,z)$. In the fourier space, they can be expressed as%
\begin{equation}
h_{tx}(t,z)=\int_{-\infty }^{\infty }\frac{d\omega }{2\pi }e^{-i\omega
t}h_{tx}(\omega ,z),\;h_{xz}(t,z)=\int_{-\infty }^{\infty }\frac{d\omega }{%
2\pi }e^{-i\omega t}h_{xz}(\omega ,z).
\end{equation}

\begin{itemize}
\item $d=2$
\end{itemize}

Let's write down the coupled EOM of two vector modes%
\begin{eqnarray}
h_{tx}^{\prime \prime }+\frac{3}{z}h_{tx}^{\prime }+i\omega h_{xz}^{\prime }+
\frac{3i}{z}h_{xz}+\frac{\alpha _{1}}{zf}h_{tx} =0, \nonumber \\
h_{xz}-\frac{iz\omega h_{tx}^{\prime }}{z\omega ^{2}+\alpha _{1}f} =0,
\label{eomd2}
\end{eqnarray}%
where the prime denotes the derivative with respect to $z$. From eq. (\ref%
{eomd2}), one can see that $h_{xz}$ is completely determined by $h_{tx}$.
Near the boundary, the asymptotic solutions read%
\begin{eqnarray}
h_{tx} =h_{tx}^{(0)}+\frac{1}{z^{1}}h_{tx}^{(1)}+\frac{1}{z^{2}}h_{tx}^{(2)}+\cdots  \nonumber \\
h_{xz} =\frac{1}{z^{3}}h_{xz}^{(3)}+\frac{1}{z^{4}}h_{xz}^{(4)}+\cdots .
\label{solud2}
\end{eqnarray}%
Here the coefficient $h_{tx}^{(0)}$ is the only independent source. Two
coefficients $h_{tx}^{(1)}$ and $h_{xz}^{(3)}$ are fixed by $h_{tx}^{(0)}$.
The exact relations are $h_{tx}^{(1)}=\alpha _{1}h_{tx}^{(0)}$ and $h_{xz}^{(3)}=-i\omega h_{tx}^{(0)}$. Other coefficients $h_{tx}^{(2)}$ and $h_{xz}^{(4)}$ rely on $h_{tx}^{(0)}$ and the incoming boundary conditions at
the horizon. Note that the presence of $h_{tx}^{(1)}$ is due to the
diffeomorphism breaking.

Expanding the on-shell action and the counterterm action above the
background, we obtain a quadratic action%
\begin{eqnarray}
S_{\mathrm{on-shell}}^{(2)}+S_{\mathrm{ct}}^{(2)} =\frac{V}{2\kappa ^{2}}\int_{-\infty }^{\infty }\frac{d\omega }{2\pi }\bigg\{-\frac{1}{4}\alpha
_{1}z^{2}\bar{h}_{xz}h_{xz}+\frac{1}{2}iz^{3}\omega \bar{h}_{tx}h_{xz}
\nonumber \\
+\left( z^{2}-\frac{z^{3}}{\sqrt{f}}\right) \bar{h}_{tx}h_{tx}+\frac{1}{2}z^{3}\bar{h}_{tx}h_{tx}^{\prime }\bigg\},  \label{qactiond2}
\end{eqnarray}%
where the modes with the bar have the argument $-\omega $. Substituting the
asymptotic solutions (\ref{solud2}) and the blackening factor (\ref{BHmetric}) into eq. (\ref{qactiond2}), we obtain the renormalized action:%
\begin{equation}
S_{\mathrm{ren}}^{(2)}=\frac{V}{2\kappa ^{2}}\int_{-\infty }^{\infty }\frac{d\omega }{2\pi } \Bigg\{\frac{1}{8}\left( \alpha _{1}^{2}-4\alpha
_{1}z_{+}-4z_{+}^{2}+4\omega ^{2}\right) \bar{h}_{tx}^{(0)}h_{tx}^{(0)}-\bar{h}_{tx}^{(0)}h_{tx}^{(2)}\Bigg\} .
\end{equation}%
One can find that it is finite.

\begin{itemize}
\item $d=3,\;\alpha _{1}=0,\;\alpha _{2}\neq 0$
\end{itemize}

For higher dimensions, our counterterms are not enough to cancel the whole
divergent part of the on-shell action in general. But when we set $\alpha
_{1}=0$ for $d=3$ or $\alpha _{1}=\alpha _{2}=0$ for $d=4$, the renormalized
action is finite. Since the derivation is similar to the previous case, we
will be a little abbreviated.

The EOM for $d=3$ are%
\begin{eqnarray}
h_{tx}^{\prime \prime }+\frac{4}{z}h_{tx}^{\prime }+i\omega h_{xz}^{\prime }+\frac{4i}{z}h_{xz}+\frac{2\alpha _{2}}{z^{2}f}h_{tx} =0,  \nonumber \\
h_{xz}-\frac{iz^{2}\omega h_{tx}^{\prime }}{z^{2}\omega ^{2}+2\alpha _{2}f} =0.  \label{eomd3}
\end{eqnarray}%
The asymptotic solutions read%
\begin{eqnarray}
h_{tx} =h_{tx}^{(0)}+\frac{1}{z^{2}}h_{tx}^{(2)}+\frac{1}{z^{3}}h_{tx}^{(3)}+\cdots   \nonumber \\
h_{xz} =\frac{1}{z^{3}}h_{xz}^{(3)}+\frac{1}{z^{5}}h_{xz}^{(4)}+\cdots ,
\label{solud3}
\end{eqnarray}%
where $h_{tx}^{(2)}=(\alpha _{2}+\omega ^{2}/2)h_{tx}^{(0)}$ and $h_{xz}^{(3)}=-i\omega h_{tx}^{(0)}$. The higher order coefficients cannot be
determined by the source $h_{tx}^{(0)}$ alone. The quadratic action can be
obtained:%
\begin{equation}
S_{\mathrm{on-shell}}^{(2)}+S_{\mathrm{ct}}^{(2)}=\frac{V}{2\kappa ^{2}}%
\int_{-\infty }^{\infty }\frac{d\omega }{2\pi }
\Bigg\{\frac{1}{2}%
iz^{4}\omega \bar{h}_{tx}h_{xz}+2\left( z^{3}-\frac{z^{4}}{\sqrt{f}}\right)
\bar{h}_{tx}h_{tx}+\frac{1}{2}z^{4}\bar{h}_{tx}h_{tx}^{\prime }\Bigg\} .
\label{qactiond3}
\end{equation}%
It follows the renormalized action%
\begin{equation}
S_{\mathrm{ren}}^{(2)}=\frac{V}{2\kappa ^{2}}\int_{-\infty }^{\infty }\frac{d\omega }{2\pi } \Bigg\{\left( z_{+}^{2}-\alpha _{2}z_{+}\right) \bar{h}%
_{tx}^{(0)}h_{tx}^{(0)}+\frac{3}{2}\left( \frac{\omega ^{2}}{2\alpha
_{2}+\omega ^{2}}-1\right) \bar{h}_{tx}^{(0)}h_{tx}^{(3)}\Bigg\} .
\end{equation}

\begin{itemize}
\item $d=4,\;\alpha _{1}=0,\;\alpha _{2}=0,\;\alpha _{3}\neq 0$
\end{itemize}

The EOM are%
\begin{eqnarray}
h_{tx}^{\prime \prime }+\frac{5}{z}h_{tx}^{\prime }+i\omega h_{xz}^{\prime }+\frac{5i}{z}h_{xz}+\frac{6\alpha _{3}}{z^{3}f}h_{tx} =0,  \nonumber \\
h_{xz}-\frac{iz^{3}\omega h_{tx}^{\prime }}{z^{3}\omega ^{2}+6\alpha _{3}f}
=0,
\end{eqnarray}%
which have the asymptotic solutions%
\begin{eqnarray}
h_{tx} =h_{tx}^{(0)}+\frac{1}{z^{2}}h_{tx}^{(2)}+\frac{1}{z^{3}}h_{tx}^{(2)}+\cdots  \nonumber \\
h_{xz} =\frac{1}{z^{3}}h_{xz}^{(3)}+\frac{1}{z^{5}}h_{xz}^{(4)}+\cdots ,
\end{eqnarray}%
with $h_{tx}^{(2)}=\omega ^{2}h_{tx}^{(0)}/2$ and $h_{xz}^{(3)}=-i\omega
h_{tx}^{(0)}$. The quadratic action is%
\begin{equation}
S_{\mathrm{on-shell}}^{(2)}+S_{\mathrm{ct}}^{(2)}=\frac{V}{2\kappa ^{2}}\int_{-\infty }^{\infty }\frac{d\omega }{2\pi }\Bigg\{\frac{1}{2}iz^{5}\omega
\bar{h}_{tx}h_{xz}+3\left( z^{4}-\frac{z^{5}}{\sqrt{f}}\right) \bar{h}_{tx}h_{tx}+\frac{1}{2}z^{5}\bar{h}_{tx}h_{tx}^{\prime }\Bigg\}.
\end{equation}%
The renormalized action is%
\begin{equation}
S_{\mathrm{ren}}^{(2)}=\frac{V}{2\kappa ^{2}}\int_{-\infty }^{\infty }\frac{d\omega }{2\pi }\Bigg\{ \left( -3\alpha _{3}z_{+}-\frac{3}{2}z_{+}^{4}+\frac{18\alpha _{3}^{3}}{\omega ^{2}}\right) \bar{h}_{tx}^{(0)}h_{tx}^{(0)}-\frac{9\alpha _{3}}{\omega ^{2}}\bar{h}_{tx}^{(0)}h_{tx}^{(3)}\Bigg\} .
\end{equation}

\section{Conclusion}

As part of the foundations of AdS/CFT correspondence, holographic
renormalization is a systematic procedure to remove the divergences by
appending the local boundary counterterms to the on-shell action. Among
several approaches to holographic renormalization, the one based on the
Hamiltonian formalism has been developed recently. The new approach starts
from the HJ equation\ and has been argued to be practical in \cite{Elvang1603}. However, it has not been clarified that whether there is a conflict with the Hamiltonian
constraint, which should be respected by any theories of gravity that are
invariant under the diffeomorphism. In this paper, we divide the HJ equation
into two parts and point out that only one part is actually used
to execute the holographic renormalization. The derivation of the CPHJ
equation does not explicitly depend on the vanishing of Hamiltonian or not,
hence being free of conflicts with the Hamiltonian constraint.

Then we apply the HJ approach to the massive gravity with different
dimensions.
Previously, by imposing the GNC and additional conditions on the
characteristic tensor of massive gravity, the standard approach was used to
build up the counterterms with $d=3$ \cite{Cao1509}. Here we only assume that the
shift vector is falling off fast enough asymptotically, indicating a little
more general situation than before. We have checked that our counterterms
are applicable at the background level. At the perturbation level, we have
shown that there are several time-dependent cases where our counterterms is
enough to cancel the divergent part of the on-shell action. Thus, our
results should be useful for the holographic calculation of thermodynamics
and transports in the strongly coupled field theories dual to massive
gravity. Moreover, we have found that the conformal anomalies appear in both
odd and even dimensions. This is different from the (pure) Einstein gravity:
it is well-known that there are no conformal anomalies in odd boundary
dimensions \cite{Henningson9806,deHaro0002}. It would be interesting to
study whether it has some profound implications on the renormalization group
flow.

Our work suggests that the HJ approach is a practical approach to
holographic renormalization, especially for the theories with conformal
anomalies. This is because the logarithmic divergences can be identified by
the same fluent procedure as the power divergences.

\acknowledgments

We thank Li-Ming Cao, Xian-Hui Ge, Wei-Jia Li, Yu Tian and Hongsheng Zhang
for helpful discussions. S. F. Wu is supported partially by NSFC with grants
No. 11675097. Y. Peng is supported in part by the National
Postdoctoral Program for Innovative Talents with Grant No. BX201700259.

\appendix

\section{Einstein-scalar theory}

We will give a brief review on the HJ approach to the holographic
renormalization of the Einstein gravity with massive scalars. More details
can be found in \cite{Elvang1603}. One can find here that the master
equation is the CPHJ equation and the procedure can be conveniently split
into three steps, which are corresponding to three subsections.

\subsection{Hamilton-Jacobi formalism}

Consider the bulk action (\ref{action}) and the ADM metric (\ref{ADM}), by
which one can obtain the Lagrangian%
\begin{eqnarray}
L &=&-\frac{1}{2\kappa ^{2}}\int_{\partial M}d^{d}x\sqrt{\gamma }N\bigg[%
R+K^{2}-K_{ij}K^{ij}-\frac{1}{N^{2}}G_{IJ}\dot{\Phi}^{I}\dot{\Phi}^{J}
\nonumber \\
&&+2\frac{N^{i}}{N^{2}}G_{IJ}\dot{\Phi}^{I}\partial _{i}\Phi ^{J}-\left(
\gamma ^{ij}+\frac{N^{i}N^{j}}{N^{2}}\right) G_{IJ}\partial _{i}\Phi
^{I}\partial _{j}\Phi ^{J}-V(\Phi )\bigg],  \label{Action2}
\end{eqnarray}%
where $K_{ij}$ is the extrinsic curvature and $R$ is the Ricci scalar on the
boundary. Then the canonical momenta conjugate to the fields can be given by%
\begin{eqnarray}
\pi ^{ij} &\equiv &\frac{\partial L}{\partial \dot{\gamma}_{ij}}=\frac{1}{%
2\kappa ^{2}}\sqrt{\gamma }(K^{ij}-K\gamma ^{ij}), \\
\pi _{I} &\equiv &\frac{\partial L}{\partial \dot{\Phi}^{I}}=\frac{1}{\kappa
^{2}N}\sqrt{\gamma }(G_{IJ}\dot{\Phi}^{J}-N^{i}G_{IJ}\partial _{i}\Phi ^{J}).
\end{eqnarray}%
Since eq. (\ref{Action2}) involves neither $\dot{N}$ nor $\dot{N}_{i}$, the
shift and lapse are Lagrangian multipliers which lead to the primary
constraints
\begin{equation}
\pi _{N}\equiv \frac{\partial L}{\partial \dot{N}}=0,\qquad \pi
_{N^{i}}\equiv \frac{\partial L}{\partial \dot{N}_{i}}=0.
\end{equation}%
The Hamiltonian can be defined by a Legendre transformation of Lagrangian
\begin{equation}
H\equiv \int_{\partial M}d^{d}x(\pi ^{ij}\dot{\gamma}_{ij}+\pi _{I}\dot{\Phi}%
^{I})-L=\int_{\partial M}d^{d}x(N\mathcal{H}+N_{i}\mathcal{H}^{i}),
\label{HMS}
\end{equation}%
where%
\begin{eqnarray}
\mathcal{H} &=&\frac{2\kappa ^{2}}{\sqrt{\gamma }}\left( \pi _{ij}\pi ^{ij}-%
\frac{1}{d-1}\pi ^{2}+\frac{1}{4}G^{IJ}\pi _{I}\pi _{J}\right) +\frac{\sqrt{%
\gamma }}{2\kappa ^{2}}\left( R-\gamma ^{ij}G_{IJ}\partial _{i}\Phi
^{I}\partial _{j}\Phi ^{J}-V(\Phi )\right) ,  \nonumber \\
\mathcal{H}^{i} &=&-2\nabla _{j}\pi ^{ij}+G_{IJ}\pi ^{I}\partial ^{i}\Phi
^{J}.
\end{eqnarray}%
An important feature of $\mathcal{H}$ and $\mathcal{H}^{i}$ is that they are
independent with $N$ and $N_{i}$. Thus, the Hamilton's equations for $N$ and
$N_{i}$ impose the secondary constraints
\begin{equation}
\mathcal{H}=0,\qquad \mathcal{H}^{i}=0,
\end{equation}%
which are called the Hamiltonian constraint and the momentum constraint,
respectively.

Furthermore, due to the diffeomorphism symmetry, one can fix the gauge%
\begin{equation}
N=1,\qquad N_{i}=0.
\end{equation}%
Then the bulk metric is simply%
\begin{equation}
ds^{2}=dr^{2}+\gamma _{ij}dx^{i}dx^{j},  \label{GNC1}
\end{equation}%
and the Hamiltonian is reduced to%
\begin{equation}
H=\int_{\partial M}d^{d}x\mathcal{H}.  \label{A01}
\end{equation}%
Consider that the canonical momenta in the Hamiltonian formalism can be
replaced by \cite{Landau1987}
\begin{equation}
\pi ^{ij}=\frac{\delta S_{\mathrm{on-shell}}}{\delta \gamma _{ij}},\qquad
\pi _{I}=\frac{\delta S_{\mathrm{on-shell}}}{\delta \Phi ^{I}}.
\label{CM D2}
\end{equation}%
One can obtain the HJ equation of Einstein-scalar theories%
\begin{equation}
H\left(\gamma _{ij},\Phi ^{I};\frac{\delta S_{\mathrm{on-shell}}}{\delta
\gamma _{ij}},\frac{\delta S_{\mathrm{on-shell}}}{\delta \Phi ^{I}}\right)+%
\frac{\partial S_{\mathrm{on-shell}}}{\partial r}=0.  \label{HJEqu}
\end{equation}

In Section 2, by decomposing the HJ equation, the CPHJ equation has been
built up%
\begin{equation}
H_{\mathrm{ct}}+\frac{\partial S_{\mathrm{ct}}}{\partial r}=0,
\end{equation}%
where $S_{\mathrm{ct}}$ denotes the (negative) divergent part of the
on-shell action and $H_{\mathrm{ct}}$ is the part of $H$ irrelevant to the
renormalized action. For later use, we rewrite $S_{\mathrm{ct}}$ in a
general form
\begin{equation}
S_{\mathrm{ct}}=-\frac{1}{\kappa ^{2}}\int_{\Sigma }d^{d}x\sqrt{\gamma }%
U(\gamma ^{ij},\Phi ^{I},r),  \label{U}
\end{equation}%
where $\Sigma $ is the hypersurface at finite radial cutoff near the
boundary. Its variation can be expressed as
\begin{eqnarray}
\frac{\delta S_{\mathrm{ct}}}{\delta \gamma _{ij}} &=&-\frac{1}{\kappa ^{2}}%
\left( \frac{1}{2}\gamma ^{ij}\sqrt{\gamma }U+\int_{\Sigma }d^{d}x\sqrt{%
\gamma }\frac{\delta U}{\delta \gamma _{ij}}\right) , \\
\frac{\delta S_{\mathrm{ct}}}{\delta \Phi ^{I}} &=&-\frac{1}{\kappa ^{2}}%
\int_{\Sigma }d^{d}x\sqrt{\gamma }\frac{\delta U}{\delta \Phi ^{I}}.
\end{eqnarray}%
Now eq. (\ref{Hnonren}) can be written by%
\begin{equation}
H_{\mathrm{ct}}=-\frac{1}{2\kappa ^{2}}\int_{\partial M}d^{d}x\sqrt{\gamma }%
\left[ \mathcal{K}+G^{IJ}P_{I}P_{J}+R-\gamma ^{ij}G_{IJ}\partial _{i}\Phi
^{I}\partial _{j}\Phi ^{J}-V(\Phi )\right] ,
\end{equation}%
where%
\begin{equation}
\mathcal{K}=4Y_{ij}Y^{ij}-\frac{1}{d-1}(U-2Y)^{2}-U^{2},  \label{EC}
\end{equation}%
\begin{equation}
Y_{ij}=\frac{\tilde{\delta}U}{\tilde{\delta}\gamma ^{ij}},\qquad Y^{ij}=-%
\frac{\tilde{\delta}U}{\tilde{\delta}\gamma _{ij}},\qquad Y=\gamma
^{ij}Y_{ij},\qquad P_{I}=\frac{\tilde{\delta}U}{\tilde{\delta}\Phi ^{I}},
\label{Yij}
\end{equation}%
and for convenience we have defined the operator:%
\begin{equation}
\frac{\tilde{\delta}}{\tilde{\delta}\mathbb{X}}\equiv \frac{1}{\sqrt{\gamma }%
}\int_{\Sigma }d^{d}x\sqrt{\gamma }\frac{\delta }{\delta \mathbb{X}}.
\label{deltab}
\end{equation}

Finally, the CPHJ equation takes the form
\begin{equation}
R+\mathcal{K}+G^{IJ}P_{I}P_{J}-\gamma ^{ij}G_{IJ}\partial _{i}\Phi
^{I}\partial _{j}\Phi ^{J}-V(\Phi )+2\frac{\partial U}{\partial r}=0,
\label{sampleHJ}
\end{equation}%
which holds as an integral equation. One can find that Eq. (\ref{sampleHJ})
is nothing but the master equation (2.15) in \cite{Elvang1603}. Here we have
shown that it should be understood as the CPHJ equation instead of the
complete HJ equation.

\subsection{Action ansatz and variation}

For simplicity, we will only involve a single massive scalar below. Then the
action is
\begin{equation}
S=-\frac{1}{2\kappa ^{2}}\int_{M}d^{d+1}x\sqrt{g}(R[g]-g^{\mu \nu }\partial
_{\mu }\Phi \partial _{\nu }\Phi -m_{\Phi }^{2}\Phi ^{2}-2\Lambda )-\frac{1}{
\kappa ^{2}}\int_{\partial M}d^{d}x\sqrt{\gamma }K.  \label{sampleAction}
\end{equation}%
Since we assume the AdS boundary, the leading asymptotic behavior of the
induced metric gives
\begin{equation}
\sqrt{\gamma }\thicksim e^{dr}\sqrt{\bar{\gamma}},
\end{equation}%
where $\bar{\gamma}_{ij}$ is the source of the boundary stress energy
tensor. This implies that the ansatz for $U$ can be organized into the
expansion
\begin{equation}
U=U_{(0)}+U_{(2)}+\cdots +U_{(2\left\lfloor \frac{d}{2}\right\rfloor )},
\end{equation}%
where $U_{(2k)}$ contains $k$ inverse metrics (or $2k$ derivatives) and $%
\left\lfloor d/2\right\rfloor $ denotes the integer no more than $d/2$. For
the Einstein-scalar theory, the potentially divergent terms in $U_{(2k)}$
are made of the scalar field $\Phi $ and boundary metric $\gamma ^{ij}$.
Using the leading asymptotic behavior of the scalar
\begin{equation}
\Phi \thicksim e^{-(d-\Delta _{\Phi })r}\bar{\Phi},
\end{equation}%
where $\Delta _{\Phi }=\frac{d}{2}+\sqrt{\frac{d^{2}}{4}+m_{\Phi }^{2}}$ is
the conformal dimension of the dual operator, one can figure out the maximal
number of the scalar that can be included in a potential divergent term. The
ansatz for the first two order is
\begin{eqnarray}
U_{(0)} &=&A_{0}(r)+A_{1}(r)\Phi +A_{2}(r)\Phi ^{2}+\cdots \\
U_{(2)} &=&B_{0}(r)R+B_{1}(r)R\Phi +B_{2}(r)R\Phi ^{2}+B_{3}(r)\Phi \square
\Phi +\cdots  \label{SU2}
\end{eqnarray}%
Note that any terms are considered as equivalent if they are related by a
total derivative. In addition, since action (\ref{sampleAction}) is
symmetric under $\Phi \leftrightarrow -\Phi $, the coefficients $A_{1}(r)$, $%
B_{1}(r)$ are simply zero.

In terms of the action ansatz, we can calculate the momenta by variations.
The relevant quantities are%
\begin{eqnarray}
\mathcal{K} &\mathcal{=}&\mathcal{K}_{(0)}+\mathcal{K}_{(2)}+\cdots \\
P_{\Phi } &=&P_{\Phi (0)}+P_{\Phi (2)}+\cdots
\end{eqnarray}%
where%
\begin{eqnarray}
\mathcal{K}_{(0)} &=&-\frac{d}{d-1}U_{(0)}^{2},\quad \mathcal{K}_{(2)}=-%
\frac{2}{d-1}U_{(0)}(U_{(2)}-2Y_{(2)})-2U_{(0)}U_{(2)}  \label{K02} \\
P_{\Phi (0)} &=&2A_{2}(r)\Phi ,\quad \quad P_{\Phi (2)}=2B_{2}(r)R\Phi
+2B_{3}(r)\square \Phi +\cdots ,  \label{P02}
\end{eqnarray}%
with%
\begin{equation}
Y_{(2)}=\gamma ^{ij}\frac{\tilde{\delta}U_{(2)}}{\tilde{\delta}\gamma ^{ij}}%
=U_{(2)}+B_{2}(r)(d-1)\square \Phi ^{2}+B_{3}(r)\left( 1-\frac{1}{2}d\right)
\nabla _{i}(\Phi \nabla ^{i}\Phi )+\cdots .  \label{Y2}
\end{equation}

\subsection{Solution of CPHJ equation}

By inserting the ansatz into the CPHJ equation (\ref{sampleHJ}) and using
the momentum-relevant quantities calculated above, one can solve the CPHJ
equation order by order. We start with the order 0 equation
\begin{equation}
-\frac{d}{d-1}U_{(0)}^{2}+P_{\Phi (0)}^{2}-m_{\Phi }^{2}\Phi ^{2}+d(d-1)+2%
\frac{\partial U_{(0)}}{\partial r}=0.
\end{equation}%
Collecting the non-functional terms, we have%
\begin{equation}
-\frac{d}{d-1}A_{0}^{2}+d(d-1)+2\dot{A_{0}}=0.
\end{equation}%
The solution is
\begin{equation}
A_{0}=-(d-1)+\mathcal{O}(e^{-dr}).
\end{equation}%
The subleading terms give only finite contribution and can be discarded
directly.\footnote{%
We notice that the integral constant happens to be at the subleading order.
Otherwise additional boundary conditions are needed to determine the
integral constant, which can complicate or even invalidate the HJ approach.
This situation is interesting and can be traced back to the fact that the
integral constant in the solution of the HJ equation is exactly an additive
constant tacked on to the on-shell action \cite{Landau1987,Goldstein2001}.}
The coefficients of $\Phi ^{2}$ can be organized into another differential
equation
\begin{equation}
\Phi ^{2}:\quad -\frac{d}{d-1}2A_{0}A_{2}+4A_{2}^{2}-m_{\Phi }^{2}+2\dot{%
A_{2}}=0.  \label{PhiSquare}
\end{equation}%
The mass of the scalar field is restricted by Breitenlohner-Freedman bound
\cite{Breitenlohner1982}%
\begin{equation}
m_{\Phi }^{2}\geq -\frac{d^{2}}{4}.
\end{equation}%
The solutions of eq. (\ref{PhiSquare}) rely on the value of mass
\begin{eqnarray}
&&m_{\Phi }^{2}=-\frac{d^{2}}{4}:\quad A_{2}=-\frac{d}{4}+\frac{1}{2r}+%
\mathcal{O}(\frac{1}{r^{2}}), \\
&&m_{\Phi }^{2}>-\frac{d^{2}}{4}:\quad A_{2}=\frac{1}{2}(\Delta _{\Phi }-d)+%
\mathcal{O}(e^{-(2\Delta _{\Phi }-d)r}).  \label{A2}
\end{eqnarray}%
We will use the solution (\ref{A2}) to continue the renormalization
procedure. Another branch is similar. Thus, $U_{(0)}$ has been specified
\begin{equation}
U_{(0)}=-(d-1)-\frac{1}{2}(d-\Delta _{\Phi })\Phi ^{2}+\cdots .
\end{equation}

We turn to the order 2 equation%
\begin{equation}
R+\left[ -\frac{2}{d-1}U_{(0)}(U_{(2)}-2Y_{(2)})-2U_{(0)}U_{(2)}\right]
+2P_{(0)}P_{(2)}-\gamma ^{ij}\partial _{i}\Phi \partial _{j}\Phi +2\frac{%
\partial U_{(2)}}{\partial r}=0.  \label{Sample2th}
\end{equation}%
Equation (\ref{Sample2th}) naturally induces the following equations%
\begin{eqnarray}
R &:&\quad 1-(4-2d)B_{0}+2\dot{B}_{0}=0, \\
R\Phi ^{2} &:&\quad \frac{(\Delta _{\Phi }-d)(2-d)}{d-1}B_{0}+(4\Delta
_{\Phi }-2d-4)B_{2}+2\dot{B}_{2}=0, \\
\Phi \square \Phi &:&\quad 1+(4\Delta _{\Phi }-2d-4)B_{3}+2\dot{B}_{3}=0.
\end{eqnarray}%
For $d=2$, the solution for $B_{0}$ is
\begin{equation}
B_{0}=-\frac{1}{2}r.
\end{equation}%
In addition, the solutions for $B_{2}$ and $B_{3}$ indicate that they are
not relevant to the divergent terms. Then%
\begin{equation}
U_{(2)}=-\frac{1}{2}rR.
\end{equation}%
For $d>2$, the solutions are%
\begin{eqnarray}
B_{0} &=&\frac{1}{4-2d}+\mathcal{O}(e^{-(d-2)r}), \\
B_{2} &=&\begin{cases} \frac{d-\Delta _{\Phi }}{4(d-1)(2\Delta _{\Phi
}-d-2)}+\mathcal{O}(e^{(d+2-2\Delta _{\Phi })r}) & \text{for }2\Delta _{\Phi
}-d-2\neq 0 \\ \frac{d-\Delta _{\Phi }}{4(d-1)}r+\mathcal{O}(1) & \text{for
}2\Delta _{\Phi }-d-2=0\end{cases}, \\
B_{3} &=&\begin{cases} -\frac{1}{2(2\Delta _{\Phi
}-d-2)}+\mathcal{O}(e^{(d+2-2\Delta _{\Phi })r}) & \text{for }2\Delta _{\Phi
}-d-2\neq 0 \\ -\frac{1}{2}r+\mathcal{O}(1) & \text{for }2\Delta _{\Phi
}-d-2=0\end{cases},
\end{eqnarray}%
which implies%
\[
U_{(2)}=%
\begin{cases} \frac{1}{4-2d}R-\frac{1}{2(2\Delta _{\Phi
}-d-2)}\left[ \Phi \square \Phi -\frac{d-\Delta _{\Phi }}{2(d-1)}R\Phi
^{2}\right] +\cdots & \text{for }2\Delta _{\Phi }-d-2\neq 0 \\
\frac{1}{4-2d}R-\frac{1}{2}\left[ \Phi \square \Phi -\frac{d-\Delta _{\Phi
}}{2(d-1)}R\Phi ^{2}\right] r+\cdots & \text{for }2\Delta _{\Phi
}-d-2=0\end{cases}
\]%
One can further deal with higher order descent equations if needed. Finally,
the counterterm action is%
\begin{equation}
S_{\mathrm{ct}}=-\frac{1}{\kappa ^{2}}\int_{\Sigma }d^{d}x\sqrt{\gamma }%
\left[ U_{(0)}+U_{(2)}+...+U_{(2\left\lfloor \frac{d}{2}\right\rfloor )}%
\right] .
\end{equation}%
This result agrees with the one given by the standard approach \cite%
{deHaro0002,Skenderis0209}.

\section{The details of computation}

Here we present the details when dealing with $\mathcal{K}$ term in the CPHJ
equation of massive gravity. The basic formulas of functional variations
with respect to the boundary metric are given in Appendix C.

\begin{itemize}
\item $Y_{(m)ij}Y_{(n)}^{ij}$
\end{itemize}

The following equations have been used when we calculate the terms $%
Y_{(m)ij}Y_{(n)}^{ij}$:%
\begin{eqnarray}
\frac{\tilde{\delta}[X^{m}]}{\tilde{\delta}\gamma _{ij}}\frac{\tilde{\delta}%
[X^{n}]}{\tilde{\delta}\gamma ^{ij}} &=&-\frac{mn}{4}[X^{m+n}],  \nonumber \\
\frac{\tilde{\delta}[X]}{\tilde{\delta}\gamma _{ij}}\frac{\tilde{\delta}R}{%
\tilde{\delta}\gamma ^{ij}} &=&-\frac{1}{2}X^{ij}R_{ij},  \nonumber \\
\frac{\tilde{\delta}[X]}{\tilde{\delta}\gamma _{ij}}\frac{\tilde{\delta}%
(X^{kl}R_{kl})}{\tilde{\delta}\gamma ^{ij}} &=&-\frac{3}{4}{R^{i}}_{j}{X^{j}}%
_{k}{X^{k}}_{i}  \nonumber \\
&&+\frac{1}{4}(2X^{ij}\nabla ^{k}\nabla _{j}X_{ki}-X^{ij}\nabla ^{k}\nabla
_{k}X_{ij}-[X]\nabla _{i}\nabla _{j}X^{ij}),  \nonumber \\
\frac{\tilde{\delta}[X]}{\tilde{\delta}\gamma _{ij}}\frac{\tilde{\delta}%
([X]R)}{\tilde{\delta}\gamma ^{ij}} &=&-\frac{1}{4}[X]^{2}R-\frac{1}{2}%
(R_{ij}X^{ij}[X]+[X]\nabla _{i}\nabla ^{i}[X]-X^{ij}\nabla _{i}\nabla
_{j}[X]).  \label{YY}
\end{eqnarray}%
Remind that the operator $\tilde{\delta}/\tilde{\delta}\gamma _{ij}$ has
been defined in eq. (\ref{deltab}).

\begin{itemize}
\item $Y_{(k)}$
\end{itemize}

One can simplify the computation by utilizing the relation between $U_{(k)}$
and $Y_{(k)}$. We state it from the beginning, when $k=1$:
\begin{equation}
Y_{(1)}=\gamma ^{ij}\frac{\tilde{\delta}U_{(1)}}{\tilde{\delta}\gamma ^{ij}}%
=\gamma ^{ij}\left( \frac{1}{2}BX_{ij}\right) =\frac{1}{2}U_{(1)}.
\label{Y1U1}
\end{equation}%
Similarly, when $k=2$,
\begin{equation}
Y_{(2)}=\gamma ^{ij}\frac{\tilde{\delta}U_{(2)}}{\tilde{\delta}\gamma ^{ij}}%
=\gamma ^{ij}(C_{1}R_{ij}+C_{2}[X^{2}]_{ij}+C_{3}[X]X_{ij})=U_{(2)}.
\label{Y2U2}
\end{equation}%
When $k=3$,
\begin{eqnarray}
&&\gamma ^{ij}\frac{\tilde{\delta}([X]R)}{\tilde{\delta}\gamma ^{ij}}=\frac{3%
}{2}[X]R+\text{total derivatives}, \\
&&\gamma ^{ij}\frac{\tilde{\delta}(R_{kl}X^{kl})}{\tilde{\delta}\gamma ^{ij}}%
=\frac{3}{2}R_{kl}X^{kl}+\text{total derivatives}, \\
&&\gamma ^{ij}\frac{\tilde{\delta}[X^{3}]}{\tilde{\delta}\gamma ^{ij}}=\frac{%
3}{2}[X^{3}],\quad \gamma ^{ij}\frac{\tilde{\delta}([X^{2}][X])}{\tilde{%
\delta}\gamma ^{ij}}=\frac{3}{2}[X^{2}][X],\quad \gamma ^{ij}\frac{\tilde{%
\delta}[X]^{3}}{\tilde{\delta}\gamma ^{ij}}=\frac{3}{2}[X]^{3}.
\end{eqnarray}%
Then one can read off
\begin{equation}
Y_{(3)}=\gamma ^{ij}\frac{\tilde{\delta}U_{(3)}}{\tilde{\delta}\gamma ^{ij}}=%
\frac{3}{2}U_{(3)}+\text{total derivatives.}  \label{Y3U3}
\end{equation}%
These total derivatives can be directly dropped after considering the fact
that the HJ equation is an integral equation where they are multiplied by
the constant $U_{(0)}$. When $k=4$, some results are exemplified:
\begin{eqnarray}
\gamma ^{ij}\frac{\tilde{\delta}(R_{kl}R^{kl})}{\tilde{\delta}\gamma ^{ij}}
&=&2R_{kl}R^{kl}+\text{total derivatives}, \\
\gamma ^{ij}\frac{\tilde{\delta}(R_{kl}{X^{l}}_{o}X^{ok})}{\tilde{\delta}%
\gamma ^{ij}} &=&2R_{kl}{X^{l}}_{o}X^{ok}+\text{total derivatives}, \\
\gamma ^{ij}\frac{\tilde{\delta}([X]\nabla _{i}\nabla ^{i}[X])}{\tilde{\delta%
}\gamma ^{ij}} &=&2[X]\nabla _{i}\nabla ^{i}[X]+\text{total derivatives} \\
&\cdots &  \nonumber
\end{eqnarray}%
This directly gives
\begin{equation}
Y_{(4)}=\gamma ^{ij}\frac{\tilde{\delta}U_{(4)}}{\tilde{\delta}\gamma ^{ij}}%
=2U_{(4)}+\text{total derivatives}.  \label{Y4U4}
\end{equation}

One can find that up to the total derivatives, the relation between $U_{(k)}$
and $Y_{(k)}$ looks like the Euler's homogeneous function theorem. It would
be interesting to give a general proof in the future.

\section{Some basic formulas}

Here we present some basic formulas that we have used. They are%
\begin{eqnarray}
\text{$\mathbb{\tilde{X}}$}\frac{\tilde{\delta}R}{\tilde{\delta}\gamma ^{ij}}%
&=&R_{ij}\mathbb{X}+(\square \mathbb{X})\gamma _{ij}-\nabla _{(i}\nabla _{j)}%
\mathbb{X},  \label{F1} \\
\text{$\mathbb{\tilde{X}}$}\frac{\tilde{\delta}(R_{kl}R^{kl})}{\tilde{\delta}
\gamma ^{ij}}&=&2R_{k(i}{R^{k}}_{j)}\mathbb{X}+\nabla _{k}\nabla _{l}(%
\mathbb{X}R^{kl})\gamma _{ij}+(\square \mathbb{X}R_{ij})-2\nabla ^{k}\nabla
_{(i}(\mathbb{X}R_{j)k}),  \label{F2} \\
\text{$\mathbb{\tilde{X}}$}\frac{\tilde{\delta}({R^{k}}_{mln})}{\tilde{%
\delta }\gamma ^{ij}}&=&\frac{1}{2}\left[ \nabla ^{k}\nabla _{l}\mathbb{X}%
\gamma _{m(i}\gamma _{j)n}-\nabla _{m}\nabla _{l}\mathbb{X}\gamma
_{n(i}\delta _{j)}^{k}-\nabla _{n}\nabla _{l}\mathbb{X}\gamma _{m(j}\delta
_{i)}^{k}\right] \!-\!(l\leftrightarrow n),  \label{F3} \\
\text{$\mathbb{\tilde{X}}$}\frac{\tilde{\delta}(\square \mathbb{Y})}{\tilde{
\delta}\gamma ^{ij}}&=&\text{$\mathbb{X}$}\nabla _{(i}\nabla _{j)}\mathbb{Y}%
-\nabla _{(i}(\text{$\mathbb{X}$}\nabla _{j)}\mathbb{Y})+\frac{1}{2}\nabla
_{k}(\text{$\mathbb{X}$}\nabla ^{k}\mathbb{Y})\gamma _{ij}+\square \text{$%
\mathbb{X}$}\frac{\delta \mathbb{Y}}{\delta \gamma ^{ij}},  \label{F4}
\end{eqnarray}%
where we have defined%
\begin{equation}
\mathbb{\tilde{X}}\frac{\tilde{\delta}}{\tilde{\delta}\gamma ^{ij}}\equiv
\frac{1}{\sqrt{\gamma }}\int_{\Sigma }d^{d}x\sqrt{\gamma }\mathbb{X}\frac{
\delta }{\delta \gamma ^{ij}},
\end{equation}%
and%
\begin{eqnarray}
\frac{\delta \lbrack X^{n}]}{\delta \gamma ^{ij}} &=&\frac{n}{2}[X^{n}]_{ij},
\label{F5} \\
X^{ij}\frac{\delta X^{kl}}{\delta \gamma ^{ij}} &=&\frac{3}{2}[X^{2}]^{kl},
\label{F6} \\
\gamma ^{ij}\frac{\delta X^{kl}}{\delta \gamma ^{ij}} &=&\frac{3}{2}X^{kl}.
\label{F7}
\end{eqnarray}%
Note that eqs. (\ref{F1})-(\ref{F4}) have originally been listed in \cite%
{Elvang1603} and eq. (\ref{F5}) was proved in \cite{Cao1511}. The rest part
of this appendix is our demonstration for eqs. (\ref{F6}) and (\ref{F7}).

\subsection{Variation of the matrix X}

The variation of a square root matrix with respect to the metric was studied
by Bernard, etc. The result is presented in (4.18) in \cite{Bernard1504},
where $S$ is $\mathcal{X}$ in our notation. Now we multiply $\mathcal{X}%
_{\rho \sigma }$ on each side of that equation and then make use of
Cayley-Hamilton theorem, which gives
\begin{equation}
\mathcal{X}_{\rho \sigma }\frac{\delta {\mathcal{X}{}^{\lambda }}_{\mu }}{%
\delta g_{\rho \sigma }}=-\frac{1}{2}{[\mathcal{X}^{2}]^{\lambda }}_{\mu }.
\end{equation}%
This equation takes the same form as (\ref{F6}). Nevertheless, (4.18) in
\cite{Bernard1504} is unsuitable to our case. The main reason is following.
The calculation in \cite{Bernard1504} is applicable only if the matrix $%
e_{3}I+e_{1}\mathcal{X}^{2}$ is invertible. In consideration of the gauge $%
f_{t\mu }=0$ that we have adopted, however, one has
\begin{equation}
e_{D-1}(\mathcal{X})=e_{D-1}(X)=det(X)=0,  \label{edx}
\end{equation}%
so $e_{3}I+e_{1}\mathcal{X}^{2}$ is actually a singular matrix when the bulk
dimension $D=4$ in our case. An applicable modification is given below. We
refer \cite{Bernard1504} for more details and notation. Let us first deal
with an even $d$. The Cayley-Hamilton theorem for the $d\times d$ matrix $X$
is given by
\begin{equation}
\sum\limits_{n=0}^{d}(-1)^{n}X^{d-n}e_{n}(X)=0.
\end{equation}%
Using (A.2) in \cite{Bernard1504} gives
\begin{equation}
\sum\limits_{n=0}^{d-1}(-1)^{n}e_{n}\delta
(X^{d-n})=\sum\limits_{n=1}^{d}\sum%
\limits_{m=1}^{n}(-1)^{n+m}e_{n-m}X^{d-n}Tr[X^{m-1}\delta X].
\end{equation}%
Here we have used $e_{0}=1$. By writing $X^{2n+1}=X\cdot X^{2n}$, one can
convert the variation of the square root matrix $X$ to the known variation $%
\delta (X^{2n})$, namely%
\begin{equation}
\frac{{(\delta X)^{k}}_{a}}{\delta \gamma ^{ij}}{\
(e_{1}X^{d-2}+e_{3}X^{d-4}+...+e_{d-1}I)^{a}}_{l}
\end{equation}%
\[
=\Big\{\Big(\delta {[X^{d}]^{k}}_{l}+e_{2}\delta {[X^{d-2}]^{k}}%
_{l}+...+e_{d-2}\delta {[X^{2}]^{k}}_{l}\Big)-{X{}^{k}}_{a}\Big(e_{1}\delta {%
\ [X^{d-2}]^{a}}_{l}+...+e_{d-3}\delta {[X^{2}]^{a}}_{l}\Big)
\]%
\[
-\sum\limits_{n=1}^{d}\Big((-1)^{n+1}e_{n-1}{[X^{d-n}]^{k}}_{l}\delta
\lbrack X]+\frac{1}{2}\sum\limits_{m=2}^{n}(-1)^{n+m}e_{n-m}{[X^{d-n}]^{k}}%
_{l}Tr[X^{m-2}\delta X^{2}]\Big)\Big\}\Big/\delta \gamma ^{ij}.
\]%
Multiplying $X^{ij}$ on both sides and working out all variation on the RHS
of this equation, one can obtain%
\begin{eqnarray}
&&X^{ij}\frac{\delta {X{}^{k}}_{a}}{\delta \gamma ^{ij}}{\
(e_{1}X^{d-2}+e_{3}X^{d-4}+...+e_{d-1}I)^{a}}_{l}  \nonumber \\
&=&\frac{1}{2}\Big\{\Big(d{[X^{d+1}]^{k}}_{l}+(d-2)e_{2}{[X^{d-1}]^{k}}%
_{l}+...+2e_{d-2}{[X^{3}]^{k}}_{l}\Big)  \nonumber \\
&&-\Big((d-2)e_{1}{[X^{d}]^{k}}_{l}+(d-4)e_{3}{[X^{d-2}]^{k}}%
_{l}+...+2e_{d-3}{[X^{4}]^{k}}_{l}\Big)  \nonumber \\
&&-\sum\limits_{n=1}^{d}\sum\limits_{m=1}^{n}(-1)^{n+m}e_{n-m}[X^{m+1}]{\
[X^{d-n}]^{k}}_{l}\Big\}.  \label{D13}
\end{eqnarray}%
Here we have used (\ref{Auxiliary1}) and (\ref{F5}). Using (\ref{Auxiliary2}%
) gives the RHS of (\ref{D13}) as
\begin{eqnarray}
&&\frac{1}{2}\Big\{\sum\limits_{n=-1}^{d-3}(-1)^{n}(n+1-d)e_{n+1}{\
[X^{d-n}]^{k}}_{l}+\Big(e_{1}{[X^{d}]^{k}}_{l}+e_{3}{[X^{d-2}]^{k}}%
_{l}+...+e_{d-3}{[X^{4}]^{k}}_{l}\Big)  \nonumber \\
&&-\Big(\sum\limits_{n=1}^{d}(-1)^{n}(n+1)e_{n+1}{[X^{d-n}]^{k}}_{l}+e_{1}{\
[X^{d}]^{k}}_{l}\Big)\Big\}  \nonumber \\
&=&\frac{1}{2}\Big\{-d\sum\limits_{n=-1}^{d-1}(-1)^{n}e_{n+1}{[X^{d-n}]^{k}}%
_{l}+\Big(e_{1}{[X^{d}]^{k}}_{l}+e_{3}{[X^{d-2}]^{k}}_{l}+...+e_{d-1}{\
[X^{2}]^{k}}_{l}\Big)\Big\}  \nonumber \\
&=&\frac{1}{2}\Big(e_{1}{[X^{d}]^{k}}_{l}+e_{3}{[X^{d-2}]^{k}}%
_{l}+...+e_{d-1}{[X^{2}]^{k}}_{l}\Big).  \label{FIresult}
\end{eqnarray}%
Here we have used $e_{n}=0$ for any $n>d$. That is
\begin{eqnarray}
&&X^{ij}\frac{\delta {X{}^{k}}_{a}}{\delta \gamma ^{ij}}{\
(e_{1}X^{d-2}+e_{3}X^{d-4}+...+e_{d-1}I)^{a}}_{l}  \nonumber \\
&=&\frac{1}{2}{[X^{2}]^{k}}_{a}{(e_{1}X^{d-2}+e_{3}X^{d-4}+...+e_{d-1}I)^{a}}%
_{l}.
\end{eqnarray}%
Whether one takes the gauge $\;f_{t\mu }=0$ or not, $%
e_{1}X^{d-2}+e_{3}X^{d-4}+...+e_{d-1}I$ is an invertible matrix commonly.
Then multiplying by its inverse on each side, one has
\begin{equation}
X^{ij}\frac{\delta {X{}^{k}}_{l}}{\delta \gamma ^{ij}}=\frac{1}{2}{\
[X^{2}]^{k}}_{l}.
\end{equation}%
This is just (\ref{F6}). The proof of an odd $d$ and (\ref{F7}) can be given
in the same way.

\subsection{Auxiliary formulas}

\begin{itemize}
\item For an even $n$, $X^{n}$ can be written as
\begin{equation}
{[X^{n}]^{k}}_{l}={[X^{2}]^{k}}_{a_{1}}\cdot {[X^{2}]^{a_{1}}}_{a_{2}}\cdots
{\ [X^{2}]^{a_{n/2-1}}}_{l}.
\end{equation}
This gives
\begin{equation}
X^{ij}\frac{\delta {[X^{n}]^{k}}_{l}}{\delta \gamma ^{ij}}=\frac{1}{2}n{\
[X^{n+1}]^{k}}_{l}.  \label{Auxiliary1}
\end{equation}

\item Equation (2.20) in \cite{Bernard1504} gives
\begin{equation}
e_{n+1}=\frac{-1}{n+1}\sum\limits_{m=1}^{n+1}(-1)^{m}[X^{m}]e_{n+1-m}=\frac{%
1 }{n+1}\Big(\lbrack X]e_{n}-\sum\limits_{m=1}^{n}(-1)^{m+1}[X^{m+1}]e_{n-m} %
\Big).
\end{equation}
That is
\begin{equation}
\sum\limits_{m=1}^{n}(-1)^{m+1}[X^{m+1}]e_{n-m}=-(n+1)e_{n+1}+e_{1}e_{n},
\end{equation}
which induces
\begin{eqnarray}
&&\sum\limits_{n=1}^{d}\sum\limits_{m=1}^{n}(-1)^{n+m}[X^{m+1}]e_{n-m}{\
[X^{d-n}]^{k}}_{l}  \nonumber \\
&=&e_{1}{[X^{d}]^{k}}_{l}+\sum\limits_{n=1}^{d}(-1)^{n}(n+1)e_{n+1}{\
[X^{d-n}]^{k}}_{l}.  \label{Auxiliary2}
\end{eqnarray}
\end{itemize}

\end{document}